%% file: ANA-TOPQ-2017-19-PAPER.tex
\PassOptionsToPackage{bookmarks=false}{hyperref}
\documentclass[cernpreprint, USenglish, subfigure, texmf]{atlasdoc}
 
\usepackage{atlaspackage}
\usepackage{atlasbiblatex}
 
\usepackage{atlasphysics}
 
\graphicspath{{logos/}{figs/}}
 
\usepackage{ANA-TOPQ-2017-19-PAPER-defs}

\AtlasTitle{Measurements of jet observables sensitive to $b$-quark fragmentation
in $t\bar{t}$ events at the LHC with the ATLAS detector}
 
\AtlasAbstract{
Several observables sensitive to the fragmentation of $b$-quarks into $b$-hadrons
are measured using 36~fb${}^{-1}$ of $\sqrt s = 13$~TeV proton--proton collision data
collected with the ATLAS detector at the LHC.
Jets containing $b$-hadrons are obtained from a sample of dileptonic $t
\bar{t}$ events, and the associated set of charged-particle tracks is
separated into those from the primary $pp$ interaction vertex and those from
the displaced $b$-decay secondary vertex.
This division is used to construct observables that characterize the
longitudinal and transverse momentum distributions of the
$b$-hadron within the jet.
The measurements have been corrected for detector effects and provide a
test of heavy-quark-fragmentation modeling at the LHC in a system where
the top-quark decay products are color-connected to the proton beam remnants.
The unfolded distributions are compared with the predictions of several modern
Monte Carlo parton-shower generators and generator tunes, and a wide range of
agreement with the data is observed, with $p$-values varying from $5 \times
10^{-4}$ to 0.98.
These measurements complement similar measurements from $e^+e^-$ collider
experiments in which the $b$-quarks originate from a color-singlet
$Z/\gamma^*$.
}
 
\author{The ATLAS Collaboration}
 
\PreprintIdNumber{CERN-EP-2021-116}

\AtlasJournalRef{Phys.Rev.D 106 (2022) 032008}
\AtlasDOI{10.1103/PhysRevD.106.032008}
\hypersetup{pdftitle={ATLAS document},pdfauthor={The ATLAS Collaboration}}
 
\begin{document}
 
\maketitle

% The next lines are included from the .//intro.tex input file
\section{Introduction}
\label{sec:intro}
 
The dominant paradigm for calculations of hadron collider physics observables
makes use of quantum chromodynamics (QCD) factorization to separate the
perturbative hard-scattering process from nonperturbative initial- and
final-state effects.
Historically, these effects were separated into three disjoint parts: a
nonperturbative structure function representing the extraction of incoming
parton configurations from the colliding hadrons, a perturbative hard process,
and a nonperturbative fragmentation function mapping the outgoing hard-process
partons to final-state hadrons and jets.
Much of early QCD physics research was conducted within this
paradigm~\cite{Field:1976ve,Feynman:1978dt}.
 
This picture has been continually refined in the quest for increased precision
and exclusive, realistic, Monte Carlo (MC) event modeling.
In particular, modeling of both the initial and final states is now typically enhanced by
perturbative analytic resummation or parton-shower algorithms, which
go beyond fixed-order in the QCD coupling to account for the enhanced effects of
multiple QCD parton emissions in singular regions of emission
phase-space~\cite{Cacciari:2001cw}. These can be considered enhancements of the
hard process, which were previously absorbed -- resulting in a loss of kinematic
accuracy -- into the parton distribution and fragmentation functions.
The partonic content of the proton, from
which the perturbative scattering is initiated, is described by the parton
distribution functions (PDFs)~\cite{Butterworth:2015oua}, while the formation of
hadrons below the hadronization scale $\sim\!\Lambda_\mathrm{QCD}$ is described
by the fragmentation functions~\cite{Andersson:1978vj,PhysRevD.27.105}.
Both are considered universal in hard-scattering processes (by virtue of the QCD
factorization theorem) and are currently obtained by fitting nonperturbative
low-scale functions to experimental data through their perturbative evolution to
the hard-scattering scales.
 
The fragmentation of $b$-quarks is particularly interesting for several
reasons.
Bottom-quarks play an important role in Large Hadron Collider (LHC)
analyses, from top quark and Higgs boson measurements to new-particle searches.
This makes it essential that the evolution and hadronization of heavy quarks be
well-understood.
As a heavy quark whose mass plays a role in
automatically regularizing the soft divergence in QCD splitting functions,
the $b$-quarks' evolution and hadronization also provides a fundamental test of perturbative QCD and the
parton-shower formalism~\cite{Mele:1990cw,MELE2017841,Nason:1999zj,Melnikov:2004bm,Cacciari:2001cw,Mele:1990yq}.
Finally, displaced weak-decay vertices make possible an unambiguous experimental
connection between observed hadrons and their ancestor parton.
Current MC event generator models~\cite{Bowler:1981sb,Andersson:1983jt}
and their tuned parameters (tunes)~\cite{Buckley:2009bj,Skands:2014pea,ATL-PHYS-PUB-2014-021} are based on $b$-fragmentation
observables measured at $e^+ e^-$ colliders such as
LEP~\cite{DELPHI:2011aa,Heister:2001jg,Abbiendi:2002vt} and
SLC~\cite{Abe:2002iq, Abe:2002iq_err}, which are experimentally and theoretically
straightforward to interpret due to the absence of QCD initial-state radiation and
an underlying event.
However, the combination of perturbative and nonperturbative effects in fragmentation
means that the more complex color flow in hadron-collider processes will affect
observables related to fragmentation~\cite{Cacciari:2002re,Corcella:2001hz}; its measurement at the LHC
complements those previously carried out at lepton colliders.
 
In this analysis, a measurement of observables sensitive to the
fragmentation of $b$-quarks at the LHC is presented, using \ttbar events observed
in the ATLAS detector in the 2015 and 2016 data-taking runs.
Choosing this process allows studies of $b$-jet fragmentation observables in
which some fraction of the $b$-quarks are color-connected to the
initial state.
The more complex QCD environment of a hadron collider may also
affect fragmentation via effects such as QCD initial-state radiation, multiple
partonic interactions (MPI), and color reconnection.

The observables used in this paper characterize the evolution of the $b$-quark momentum into the
lower momentum of the weakly decaying $b$-hadron.
This is achieved by comparing the momenta of reconstructed $b$-hadrons with their
corresponding $b$-jets and with the event-wide momentum scale set by other
top-quark decay products.
While there is an immediate theoretical objection to making a one-to-one identification between
quarks and jets -- a colored quark cannot in isolation hadronize into a color-neutral
jet -- observables inspired by this simple picture can be well defined, and
receive finite and calculable corrections from the initiator's color
connections~\cite{Dokshitzer:1995zt}.
In the \elecpos\ experiments where such observables were previously
measured, incoming beam particles have a well-defined momentum, and hence there
is no question as to the energy of the resulting $b$-quarks in a two-jet
$\elecpos \to b\bar{b}$ process: at leading order, each must have the same energy as the incoming
electron or positron beam.
The fragmentation can hence be probed via the simple function
$x_B = E_B / E_\text{beam}$
-- the energy of the $b$-hadron divided by that of the
beam.
Due to the nature of hadronic initial states, the partonic center-of-mass or
energy scale is not known \emph{a priori} for any particular collision event.
For this reason, observables are defined using only the final-state particles
rather than the beam energy.
 
It is experimentally challenging to determine whether electrically neutral
particles originated from heavy hadron decays, the fragmentation products of the
jet, the underlying event, or pileup interactions; as such, this study focuses
on the charged constituents of the jets.
Charged-particle tracks have better momentum and angular resolution than
calorimeter clusters at the energies involved in this analysis and allow for the
removal of contamination particles emitted from pileup interactions.
To reconstruct these $b$-hadrons, inner-detector tracks are used to build
secondary vertices displaced from the hard-scatter
collision~\cite{ATL-PHYS-PUB-2017-011}.
The constituents of secondary vertices correspond to the stable, charged
$b$-hadron descendants.
The $b$-hadron's ``charged momentum'' is then calculated as the vectorial sum of
these descendants.
The $b$-hadron charged momentum is compared with various quantities that provide
an energy scale for the surrounding environment.
 
One choice is to relate the $b$-hadron with the surrounding jet of particles,
constructed via standard clustering techniques~\cite{Salam_2010}.
This is natural since the
aim of all jet algorithms is to cluster flows of energy corresponding
to well-separated initial partons in a theoretically safe way: a jet containing
a $b$-hadron is a reasonably well-defined proxy for its initiating $b$-quark.
Observables based on the ratio of $b$-hadron to $b$-jet momenta are by their
nature most sensitive to the small-angle gluon emissions from the $b$-quark
before hadronization, and therefore provide a view of $b$-quark fragmentation
that is strongly complementary to that of the lepton-collider measurements,
which rely on event-wide -- rather than local -- hadronic energy flow.
Since the $b$-hadrons are reconstructed and identified through secondary
vertices built from charged-particle tracks, the jet momentum is built from
reconstructible charged constituents, putting the $b$-hadron and the jet on the
same footing and allowing the $b$-hadron's energy to be distinguished from the
energy of the nearby hadronic activity.
Two observables are measured in this vein: (1) the ratio of the $b$-hadron
and jet charged momenta transverse to the incoming proton beams (\zbt) and (2)
the fraction of the jet charged momentum carried by the $b$-hadron decay
products along the direction of the jet charged momentum (\zbl).
The analysis object-selection methodology is therefore focused on the identification
of charged particles from primary and secondary vertices.

A comparison between the kinematic properties of the $b$-hadron and the \ttbar\
system is also made.
This measurement is performed in $e\mu \nu\nu bb$ events, and while the
primary role of the electron and muon is to select a pure sample of \ttbar\
events, the lepton momenta are correlated with the momenta of the
top quarks from which they were produced. Although the leptons are far from being a
perfect proxy for the top quarks, their momenta can be determined more precisely
than those of the $b$-hadrons and therefore can be used to characterize the
$b$-hadrons in the context of the \ttbar\ system.
The observable $\rho$ is defined as the ratio of the transverse
momentum of the $b$-hadron to the average transverse momentum of the two
leptons in the event, where ``transverse'' is defined relative to the colliding
proton beams.
As opposed to \zbt and \zbl, $\rho$ is sensitive to radiation emitted in the
top-quark decay, regardless of whether or not it was contained in the $b$-jet.

In addition to the above observables that focus on the $b$-hadron momentum,
the number of stable, charged decay products of the $b$-hadron (\nsvtrk) is
measured.
This observable is sensitive to the modelling of $b$-hadron production and
decay.

For independence from detector biases, and for maximum impact on the development of
both analytic and MC generator $b$-quark fragmentation modeling, the results
have been corrected to a fiducial acceptance definition at stable-particle
level. Particles are considered stable if they have a mean lifetime $\tau >
33$~ps ($c \tau > 10$~mm).
The unfolded data are then compared with predictions from several commonly-used
MC models and tunes, which are largely based on $e^+e^-$ $x_B$ distributions.

The organization of this paper is as follows. Section~\ref{sec:detector} gives an
overview of the ATLAS detector, and Section~\ref{sec:mcsamples} outlines the
collision data and MC simulation used in this measurement.
Section~\ref{sec:analysis} describes the object and event selection used in the
experimental analysis as well as the fiducial definition to which the observed
data are unfolded.
The unfolding procedure and systematic uncertainties are presented in
Sections~\ref{sec:unfolding} and~\ref{sec:systematics}, respectively.
Finally, the unfolded data are shown in Section~\ref{sec:results} and compared
with a number of modern MC generators tuned to \elecpos\ collider data.

% End of text imported from the .//intro.tex input file
 
% The next lines are included from the .//detector.tex input file
\section{ATLAS detector}
\label{sec:detector}
 
The ATLAS detector~\cite{PERF-2007-01,ATLAS-TDR-19,PIX-2018-001} is a multipurpose particle detector with
a forward/backward-symmetric cylindrical geometry. The detector has a nearly
$4\pi$ coverage in solid angle\footnote{ATLAS uses a right-handed coordinate
system with its origin at the nominal interaction point~(IP) in the center of
the detector and the $z$-axis along the beam pipe. The $x$-axis points from
the IP to the center of the LHC ring, and the $y$-axis points
upward. Cylindrical coordinates $(r, \phi)$ are used in the transverse plane,
$\phi$ being the azimuthal angle around the beam pipe.  The pseudorapidity is
defined in terms of the polar angle $\theta$ as $\eta = -\ln \tan(\theta/2)$, and
the rapidity is defined as $y = (1/2)\ln[(E+p_z)/(E-p_z)]$. Angular distances are
defined as $\Delta R = \sqrt{(\Delta\eta)^2 + (\Delta\phi)^2}$ and
$\Delta R_y = \sqrt{(\Delta y)^2 + (\Delta\phi)^2}$.}
and consists of an inner tracking detector, electromagnetic and hadronic
calorimeters, and a muon spectrometer. The inner detector is surrounded by a
superconducting solenoid providing a 2T magnetic field and covers a
pseudorapidity range of $|\eta| < 2.5$. The inner detector is composed of
silicon pixel and microstrip detectors as well as a transition radiation
tracker. The innermost pixel detector layer, called \textit{insertable $b$-layer} (IBL)
was installed before the start of Run~II.
The high-granularity lead/liquid-argon~(LAr) electromagnetic sampling
calorimeters measure electromagnetic energies in the
pseudorapidity region $|\eta| < 3.2$. Hadron energies are measured by a
hadronic, steel/scintillator tile calorimeter with $|\eta| < 1.7$. The forward
and endcap regions with $1.5 < |\eta| < 4.9$ are instrumented with LAr
calorimeters for both the electromagnetic and hadronic measurements. Surrounding the
calorimeters, the muon spectrometer consists of three large superconducting
toroids with eight coils each. The muon spectrometer has a system of precision
tracking chambers covering $|\eta| < 2.7$, consisting of monitored drift tubes
and, in the forward region, cathode-strip chambers. In addition, it has
fast trigger chambers covering $|\eta| < 2.4$, consisting of resistive-plate chambers in the barrel region
and thin-gap chambers in the endcaps. A two-level trigger system is used to select the events
that are recorded~\cite{TRIG-2016-01}.  The first-level trigger~(L1) is implemented in
hardware and utilizes partial detector information to accept events at a
rate below 100~kHz from the 40~MHz bunch crossings. The high-level
trigger~(HLT) is software-based and accepts events at a rate of 1~kHz.
% End of text imported from the .//detector.tex input file
 
% The next lines are included from the .//mcsamples.tex input file
\section{Data sample and simulation}
\label{sec:mcsamples}
 
The measurements in this analysis are based on data collected by the ATLAS
detector in 2015 and 2016 at a center-of-mass energy of $\sqrt{s}=13~\TeV$.
The recorded data correspond to an integrated luminosity of approximately
36.1~fb$^{-1}$. Collision events are analyzed only if they satisfy the
data quality criteria~\cite{DAPR-2018-01}, the beam conditions were stable when
they were recorded, and all subdetectors passed the requirements for operational
status.
An extensive software suite~\cite{ATL-SOFT-PUB-2021-003} is used in the
reconstruction and analysis of real and simulated data, in detector operations,
and in the trigger and data acquisition systems of the experiment.
 
This analysis requires at least one lepton to be identified in the trigger
system for each candidate event.
Depending on the data-taking period, different single-lepton trigger thresholds
were used for electrons and muons~\cite{Aad:2019wsl,Aad:2020uyd} in order to
maintain readout rates within specification over changes to the LHC
instantaneous luminosity.
For 2015 data, single-electron triggers with $\pt$ thresholds of 24~\GeV,
60~\GeV\ and 120~\GeV\ and single-muon triggers with $\pt$ thresholds of
20~\GeV\ and 50~\GeV\ were used.
For 2016 data, the $\pt$ thresholds were increased to 26~\GeV, 60~\GeV\ and 140~\GeV\
for electrons and 26~\GeV\ and 50~\GeV\ for muons.
The triggers with the
with the lowest $p_T$ thresholds include isolation requirements, while for the
triggers with higher thresholds those requirements are relaxed to increase the
acceptance.
 
Several simulated event samples are used in estimating the Standard
Model prediction for this measurement, particularly in obtaining the
corrections for detector effects discussed in Section~\ref{sec:unfolding}.
The \ttbar signal process is simulated at
next-to-leading order~(NLO) in perturbative QCD using the HVQ program~\cite{Frixione:2007nw,frixione2007powheghvq}
implemented in the \POWHEGBOX[v2]~\cite{Nason:2004rx,Frixione:2007vw} event generator with the NNPDF3.0
PDF sets~\cite{Ball:2014uwa}.
Parton-level configurations obtained with \POWHEGBOX
are processed with the \PYTHIA[8.230]~\cite{Sjostrand:2014zea} generator to model the parton shower (PS),
hadronization and underlying event, using the A14 set of tuned
parameters~\cite{ATL-PHYS-PUB-2014-021} and the NNPDF2.3 PDF set.
The A14 tune of \PYTHIA[8] is a tune of PS and MPI parameters that leaves the hadronization
parameters at their default values and uses the Lund--Bowler fragmentation model~\cite{Bowler:1981sb}.
The top-quark mass, \mt is set to 172.5~\GeV\ in all simulated samples.
The \hdamp parameter, which controls the transverse momentum of the first additional gluon emission
beyond the Born configuration, is set equal to 1.5 times the mass of the top
quark.
The main effect of this choice is to regulate the high-\pT\ emission
against which the \ttbar system recoils.
The factorization and renormalization
scales are set to $\sqrt{\mt^2 + \pt^2}$, where
\pt\ is the transverse momentum of the top quark.
 
The \ttbar\ cross-section is normalized to
$\sigma_{\ttbar}=832^{+46}_{-51}$~pb, as calculated at next-to-next-to-leading
order~(NNLO) with next-to-next-to-leading logarithmic soft-gluon
terms~\cite{CAC-1101,PRL-109-132001,JHEP-1212,JHEP-1301,Czakon:2013goa} using
the \textsc{Top++}~2.0 program~\cite{CZA-1101}.
The PDF- and \alphas-induced
uncertainties in this cross-section are calculated using the PDF4LHC
prescription~\cite{PDF4LHC} with the \MSTW[nnlo] 68\%~CL PDF~\cite{MAR-0901,MAR-0902},
\CT[10nnlo] PDF~\cite{Lai:2010vv,CT10NNLO} and
NNPDF2.3\,5f~FEN PDF~\cite{Ball:2012cx} PDF and are added in quadrature with the
uncertainties obtained from the independent variation of the factorization and
renormalization scales.
 
In order to evaluate the signal modeling uncertainties, alternative \ttbar\ samples were produced.
To evaluate the impact of initial-state radiation (ISR), two additional
\POWPY[8] samples were generated.
In one sample, the factorization and
renormalization scales in the matrix element are scaled down by a factor of two,
while at the same time setting $\hdamp = 3\mt$ and using the Var3c~Up variation
of the A14 tune.
The other sample uses factorization and
renormalization scales increased by a factor of two, while keeping \hdamp at the
nominal value of~1.5\mt and using the Var3c~Down variation.
The Var3c tune variations
correspond to a variation of \alphas in the initial state.
To account for
final-state radiation (FSR) uncertainties, two \POWPY[8] samples with the Var2~Up
and~Down variations of the A14 tune are used.
For Var2, the values of both \alphas in the final state
and parameters sensitive to ISR are varied.
A sample with \POWHEGBOX
interfaced with \HERWIG[7.04]~\cite{Bahr:2008pv,Bellm:2015jjp} with the H7UE
tune~\cite{Bellm:2015jjp} and the \MMHT[lo68cl]~\cite{Harland-Lang:2014zoa} PDF~set is used to probe the
impact of using a different parton shower and hadronization model.
 
Single-top-quark production in the $tW$ channel is simulated using the
\POWHEGBOX[v1] event generator with the CT10 PDF sets.
It is interfaced with the \PYTHIA[6.428] generator~\cite{Sjostrand:2006za} to
model the parton shower and hadronization,  using the Perugia2012 set of tuned
parameters~\cite{Skands:2010ak} and the CTEQ6L1 PDF sets~\cite{Pumplin2002}.
The total cross-section for $tW$ production
is calculated at NLO with NNLL soft-gluon corrections~\cite{Kidonakis:2015nna}.
The interference between \ttbar and $tW$ production leads to
an ambiguity in the definition of these processes
starting at NLO.
This ambiguity is removed from the $tW$ sample by
via the diagram-removal scheme~\cite{Frixione:2008yi}.
An alternative $tW$ sample with the diagram-subtraction scheme is also used.
 
The $Z/\gamma^*$+jets process is simulated using \SHERPA[2.2.1]~\cite{Gleisberg:2008ta} with the
NNPDF3.0 PDF set.
The matrix elements are calculated using
Comix~\cite{Gleisberg:2008fv} and OpenLoops~\cite{Cascioli:2011va}, for up to
two parton at~NLO and up to four partons at leading order~(LO) in the QCD
coupling.
The \MEPSatNLO prescription~\cite{Hoeche:2012yf} is used to merge the
matrix element and the parton shower contributions.
The total cross-section is normalized to the
NNLO calculation~\cite{Gavin:2010az}.
Electroweak diboson processes are
simulated using \SHERPA[2.1.1] with the CT10 PDF sets;
The matrix elements are calculated using Comix and OpenLoops.
The matrix elements for the $WW$ and $WZ$ processes are calculated with no
additional partons at NLO.
Matrix elements for the $ZZ$ process are calculated with up to one additional parton at NLO.
The matrix-element calculations for all three diboson processes ($WW$, $WZ$, and
$ZZ$) are performed for up to three additional partons at LO.
 
All nominal simulated samples are processed through a simulation
of the ATLAS detector~\cite{SOFT-2010-01} implemented using the \GEANT~\cite{Agostinelli:2002hh}
framework.
A ``fast simulation'', using parameterized showers in the
calorimeter-interaction modeling~\cite{ATL-PHYS-PUB-2010-013}, is used for
\ttbar samples with variations modeling systematic uncertainties.
Additional inclusive $pp$
interaction events generated using \PYTHIA[8.186]~\cite{Sjostrand:2007gs}
and the A3 set of tuned parameters~\cite{ATL-PHYS-PUB-2016-017} are
overlaid on all simulated signal and background events to simulate the effect of
multiple $pp$ interactions in each bunch crossing (pileup).
For all samples except the ones generated with \SHERPA{}, the decays of $b$- and
$c$-hadrons are modeled using the \EVTGEN[1.6.0] program~\cite{Lange:2001uf}.
 
The same reconstruction algorithms and analysis procedures are applied to both
data and MC simulation, allowing direct comparisons of the respective
reconstruction-level quantities and thus extraction of detector-effect corrections.
% End of text imported from the .//mcsamples.tex input file
 
% The next lines are included from the .//analysis.tex input file
\section{Analysis}
\label{sec:analysis}
 
To obtain a high-purity sample of reconstructed \bjets, this analysis uses
dilepton \ttbar ($e\mu \nu\nu bb$) events, which provide a high cross-section source of \bjets with
relatively low contamination expected from light-flavor jets. The $e\mu$ channel is used
exclusively, as it has particularly small contributions from non-\ttbar
processes.
Events with an electron, a muon, and exactly two reconstructed jets are selected.
 
Biases in reconstruction-level observables introduced by the detector and
reconstruction algorithms are corrected to a fiducial particle-level event and
object selection, designed to closely match the procedure followed for data
events.
Systematic uncertainties due to possible discrepancies between the observed and predicted
data are taken into account by introducing variations of the
predictions for background processes and of the response matrices for the \ttbar
signal; the considered sources of systematic uncertainty are discussed in
Section~\ref{sec:systematics}.
These are propagated through a fully Bayesian unfolding
(FBU) procedure~\cite{Choudalakis:2012hz} into the final unfolded distributions,
as described in Sections~\ref{sec:unfolding} and~\ref{sec:systematics}.
 
\newpage
 
\subsection{Object definitions and selection}
\label{ssec:selection}

\paragraph{Leptons:}
To reduce the number of background events with nonprompt leptons, \textit{tight} identification
criteria for the reconstructed electrons and \textit{medium} criteria for the muons are
imposed~\cite{PERF-2017-01,Aad:2016jkr,ATLAS-CONF-2020-030}.
Electrons are additionally required to fall within the fiducial volume of the
electromagnetic calorimeter ($|\eta| < 2.47$, excluding $1.37 < |\eta| <
1.52$), and muons must be central ($|\eta| < 2.5$).
The invariant mass of the two leptons, $m_{\ell \ell}$, has to be larger than 15~\GeV.
At least one selected lepton has to be matched to a lepton that fulfilled one of the
trigger decisions discussed in Section~\ref{sec:mcsamples}. Both the electrons and muons must
have $\pt > 25~\GeV$ to avoid large trigger-efficiency
uncertainties for leptons with low \pt, and be isolated from significant energy
deposits in the calorimeters and from high-momentum tracks.
For 2016 data, at least one lepton with $\pt > 27~\GeV$ is required in order to account for the higher
trigger threshold.
Differences in reconstruction and identification efficiencies between data and
simulated events are accounted for by applying scale factors derived in
$Z\rightarrow \ell\ell$ events, and the lepton energy and momentum scale and
resolutions are calibrated in data by inspecting the di-lepton mass
spectrum near to the $Z$ peak~\cite{PERF-2017-01,Aad:2016jkr}.

The particle-level fiducial lepton definition requirements are $\pt > 25~\GeV$
and $|\eta| < 2.5$, and the leptons must not have final-state hadron ancestors
(i.e.\ they are ``prompt'').
The four-momenta of photons that do not originate from hadron decays and are
within a cone of size $\Delta{R} = 0.1$ around the lepton direction are added to
the lepton four-momentum.
This is later referred to as ``photon dressing'', and produces ``dressed
leptons''.

\paragraph{Jets:}
Detector-level jets are built from topological clusters of energy deposits in
calorimeter cells calibrated to the electromagnetic interaction scale, using the
anti-$k_t$ algorithm with a radius parameter of $R=0.4$~\cite{Cacciari:2008gp}
implemented in FastJet~\cite{Fastjet}.
The resulting jets' transverse momenta are further corrected to the
corresponding particle-level jet \pt using simulation and
in situ methods~\cite{JETM-2018-05}.
After these calibrations, jets with $\pt < 30~\GeV$ or $|\eta| > 2.5$ are
removed.
A multivariate discriminant method (JVT)~\cite{PERF-2014-03} is used to remove
jets with $\pt < 60~\GeV$ and $|\eta| < 2.4$ that have large estimated energy
fractions from pileup collision vertices; above this \pt scale, this
requirement is unnecessary.
Simulated jets are corrected for JVT efficiency differences relative to the
collision data. 
 
In order to avoid double-counting of energy deposits from leptons in
calorimeter jets, those jets that are reconstructed very near to selected
leptons, with $\Delta R_y (\text{jet}, \text{lepton}) < 0.2$, are removed
because they are the result of calorimeter deposits from the leptons themselves,
where $\Delta R_y (\text{jet}, \text{lepton}) = \sqrt{\Delta \phi ^2 + \Delta
y^2}$.
On the other hand, leptons that are reconstructed within the range $0.2 < \Delta
R_y (\text{jet}, \text{lepton}) < 0.4$ are known to mostly
come from heavy-hadron decays and are not considered prompt-lepton
candidates\cite{PERF-2017-01,Aad:2016jkr,ATLAS-CONF-2020-030}.

\paragraph{Charged-particle tracks and secondary vertexing:}
 
Reconstructed tracks are limited to the fiducial volume of the inner tracker,
$|\eta| < 2.5$, and are required to have $\pt > 500~\MeV$.
Two track-identification
working points are used: {\textit \tightprimary} and {\textit \tightsecondary}.
The \tightprimary working point is optimized to achieve a high efficiency for
tracks originating from the $pp$ collision vertex while rejecting tracks
originating from random crossings of charged particles from pileup
vertices~\cite{ATL-PHYS-PUB-2015-051}; \tightprimary tracks must be
constituents of the reconstructed hard-scatter vertex.
In order to achieve a higher efficiency for charged particles from $b$-hadron
decays, \tightsecondary tracks differ from \tightprimary tracks in two main
ways:
(1)  \tightsecondary tracks are not required to be constituents of the
hard-scatter vertex but must have a transverse impact parameter satisfying $|d_0| < 3.5$~mm,
and (2) while \tightprimary tracks must have at least nine hits in the silicon pixel or microstrip layers, including at least one
insertible B-layer (IBL) or B-layer pixel hit, \tightsecondary tracks need only seven
silicon hits and are not required to have an IBL or B-layer
hit~\cite{ATL-PHYS-PUB-2017-011,ATLAS:2020jip,Aad:2015ydr}.

The measured observables require identification of the charged decay products of
$b$-hadron candidates, which are reconstructed from inner-detector tracks
matched to calorimeter jets by identifying displaced secondary vertices.
The matching of detector-level tracks to jets uses a cone whose angular radius
shrinks with increasing jet \pt from $\Delta R = 0.42$ for 30~\GeV\ jets to an asymptotic
minimum of $\Delta R = 0.24$:
 
\begin{equation*}
\Delta R_\mathrm{max} = 0.24 + \mathrm{e}^{- [1.22 + (1.64 \times 10^{-3}) \pt / \GeV]}.
\end{equation*}
 
This matching was optimized to maximize discrimination between jets that
contain $b$-hadrons and those that do not.
Secondary vertex (SV) construction from these
\tightsecondary tracks is performed with the
Single Secondary Vertex Finder algorithm~\cite{ATL-PHYS-PUB-2017-011};
secondary vertices are required to have
at least three associated tracks, since those with two or fewer were found to be
dominated by cases in which at least one track was either not reconstructed or
not properly matched to the secondary vertex.
 
The jet charged momentum is constructed using tracks from the
hard-scatter primary interaction vertex (PV)~\cite{Aaboud:2018xwy}
in addition to those from the $b$-hadron candidate secondary vertex.
PV tracks are matched to calorimeter jets via
ghost-association~\cite{Cacciari:2008gn}; in this procedure, tracks are used as
infinitesimal-energy (``ghost'') inputs to jet reconstruction, and each track is
considered matched to the jet of which its ghost is a constituent.
The total jet charged momentum is built from the union of PV and SV tracks.

\paragraph{$b$-tagging}
Jets that are likely to contain a $b$-hadron are tagged using a multivariate
discrimination algorithm (\textsc{MV2c10}) based on inputs from charged-track
impact parameter and secondary vertex reconstruction with respect to the
hard-scatter PV.
The hard-scatter PV is selected as the one with at least two associated tracks
and the highest $\sum \pt^2$, where the sum extends over all tracks with $\pt >
400$~\MeV\ matched to the vertex.
The tagger performance was optimized using $b$-jets, $c$-jets, and light-flavor jets in
\ttbar events to maintain a high tagging efficiency for \bjets while maximally
rejecting lighter-flavor jets.
A working point with 70\% \bjet tagging efficiency with an approximate light-flavor jet
(charm-jet) rejection factor of 400 (10) is used.
 
At particle level, jets are formed by clustering all final-state particles within the
calorimeter acceptance $|\eta|<4.9$, except neutrinos and prompt electrons and
muons.
Like detector-level jets, they are built using the anti-$k_t$ algorithm with
$R=0.4$.
Fiducial jets are required to have $\pt > 30~\GeV$ and $|\eta| < 2.5$.
Weakly decaying $b$-hadrons with $\pt > 5~\GeV$ are matched to jets by
finding the closest jet within $\Delta R < 0.3$ to each $b$-hadron.
The jets are then labeled as having $b$-quark flavor if they have exactly one
associated weakly decaying $b$-hadron.
Jets with two associated $b$-hadrons are explicitly excluded since they have very
different features; the fraction of such jets predicted in
simulation was found to be below the per-mille level.

\subsection{Detector event and probe-jet selection}
 
In the detector-level selection, at least one lepton is required to be
identified in the trigger system for each candidate event, using a logical
\textsc{or} of the most inclusive electron and muon triggers.
Scale factors are applied to the simulation in order to correct for known differences in trigger
efficiencies between the simulation and collision data~\cite{TRIG-2016-01}, with
residual uncertainties accounted for as described in Section~\ref{sssec:detectoruncerts}.
 
In order to reject backgrounds with fewer than two prompt leptons
(e.g.~multijet and \wjets events), exactly two reconstructed leptons with
opposite charge are required.
Contributions from backgrounds with $Z$ bosons are suppressed by requiring
one lepton to be an electron and the other to be a muon.
Exactly two reconstructed jets (after the lepton overlap removal is applied) are
required in order to reject \ttbar events with light-flavor jets from initial-
or final-state radiation.
These two jets must be separated by $\Delta R > 0.5$ to avoid $b$-hadron decay
products being matched to a jet that does not contain the initial $b$-hadron.
 
A ``tag-and-probe'' technique, described below, is employed to measure the
observables of interest for the selected jets.
If the leading jet in the event is $b$-tagged and the subleading jet has both
$|\eta| < 2.1$ and a
reconstructed secondary vertex with at least three associated tracks, then the
subleading jet is considered a ``probe-jet'' and its observables
are measured.
The more stringent $\eta$ requirement for probe-jets ensures that the full jet
area falls inside the ATLAS tracker volume.
This process is then repeated with the roles of the leading and subleading jets
reversed, and the fragmentation observables for all probe-jets are measured.
Using this procedure it is possible for both jets, only one jet, or even neither
jet to be measured, but care is taken to not bias the measurement.
After an event preselection requiring exactly two jets and at least one
\btag{}ged jet, both jets are valid probe-jets in about 25\% of \ttbar\ events,
and only one jet is a valid probe-jet about 17\% of the time.
There is a small correlation between the jets' probability of being
measured, but this correlation is well-modeled by the simulation.

\subsection{Particle-level fiducial event selection and observables}
\label{ssec:observables}
 
For the fiducial event selection, particle-level \ttbar\ events must contain
exactly one prompt electron and one prompt muon, each with $|\eta| <
2.5$ and $\pt > 25~\GeV$ after photon dressing.
The leptons must have opposite charge.
In addition, exactly two $b$-tagged jets with
$|\eta| < 2.1$, $\pt > 30~\GeV$, and $\Delta R (\text{jet}, \text{jet})> 0.5$
are required. If a particle-level \ttbar event passes the above selection, both
jets are considered as possible fiducial probe-jets.

Fiducial observables reflect their detector-level counterparts as closely as
possible to reduce dependence on the simulation during unfolding. They are
calculated using stable, charged particles with $\pt > 500~\MeV$, which are called
``fiducial charged particles''.
Each fiducial probe-jet must contain exactly one weakly decaying $b$-hadron with
$\pt > 5~\GeV$ and at least three fiducial-charged-particle decay products.
The charged momentum of the $b$-hadron, \pbcvec, is then defined as the momentum sum of all
fiducial charged particles from the $b$-hadron decay, and the $b$-jet charged
component, \pjcvec, is built from all fiducial charged particles that either are
jet constituents or originate from the $b$-hadron decay.
The dressed leptons in the event are used to calculate the value of the $\rho$ observable
defined in Section~\ref{sec:intro} and again below.

To summarize, the four observables measured are the ratio of each $b$-hadron
and jet charged momenta transverse to the incoming proton beams,
\begin{align*}
\zbtc &= \frac{\ptbc}{\ptjc} \,\,\, ,
\end{align*}
 
the fraction of the jet charged momentum carried by the $b$-hadron decay
products along the direction of the jet charged momentum,
\begin{align*}
\zblc &= \frac{\pbcvec \cdot \pjcvec}{|\pjc|^2} \,\,\, ,
\end{align*}
 
the ratio of the transverse momentum of the $b$-hadron to the average transverse
momentum of the two leptons in the event,
\begin{align*}
\rho &= \frac{2\ptbc}{\pt^e + \pt^\mu} \,\,\, ,
\end{align*}
 
and the number of fiducial charged particles produced in the $b$-hadron decay,
\nsvtrk.

\subsection{Sources of background}
 
Several sources of background jets are expected in 13~\TeV\ $pp$ collisions. A
``background probe-jet'' here is defined as a jet which passes the
detector-level requirements but fails the fiducial definition outlined in
Sections~\ref{ssec:selection} and~\ref{ssec:observables}.
Such jets can originate from \ttbar $\to e\mu b\bar{b}$ events in which one
\bjet from a top-quark decay lies outside the detector acceptance but another
jet (often from initial- or final-state radiation) is reconstructed as a valid
detector-level probe.
They can also arise from \ttbar $\to e\mu b\bar{b}$ events that fail
the fiducial event definition but pass the detector-level cuts.
Both sources of \ttbar induced backgrounds are estimated using MC simulation.
 
Non-\ttbar processes with at least one prompt electron, at least one prompt
muon, and jets are another source of background probe-jets and are also
evaluated with MC simulation.
In this analysis, the $tW$, Drell--Yan, and diboson production processes are
identified as non-\ttbar backgrounds with prompt leptons.
 
Finally, backgrounds with fewer than two prompt leptons (dominated by \ttbar and
single-top-quark production) are expected to account for less than 1\% of selected
events and are estimated from the MC simulation.

\subsection{Comparison between prediction and observed data}
 
\begin{figure}[h!]
\centering
 
\subfigure[][]{
\includegraphics[width=0.45\textwidth]{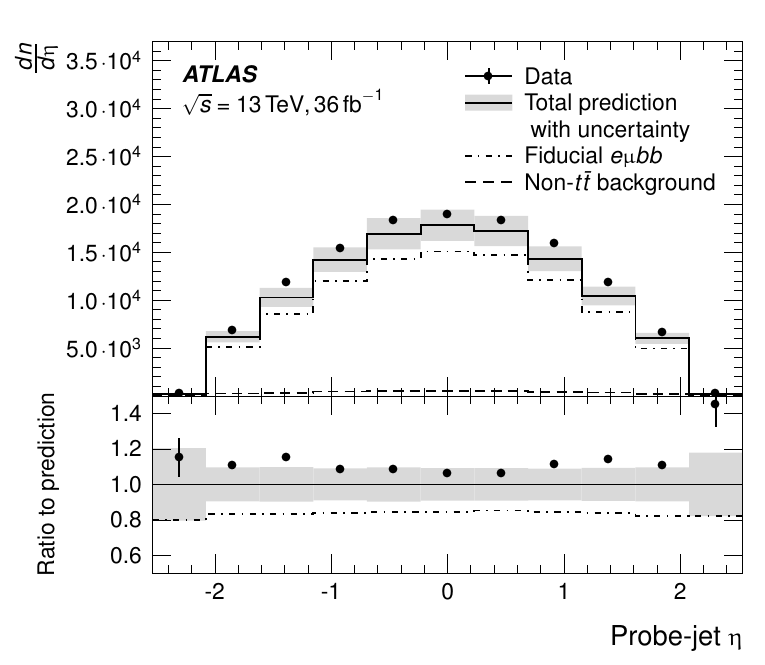}
}
\subfigure[][]{
\includegraphics[width=0.45\textwidth]{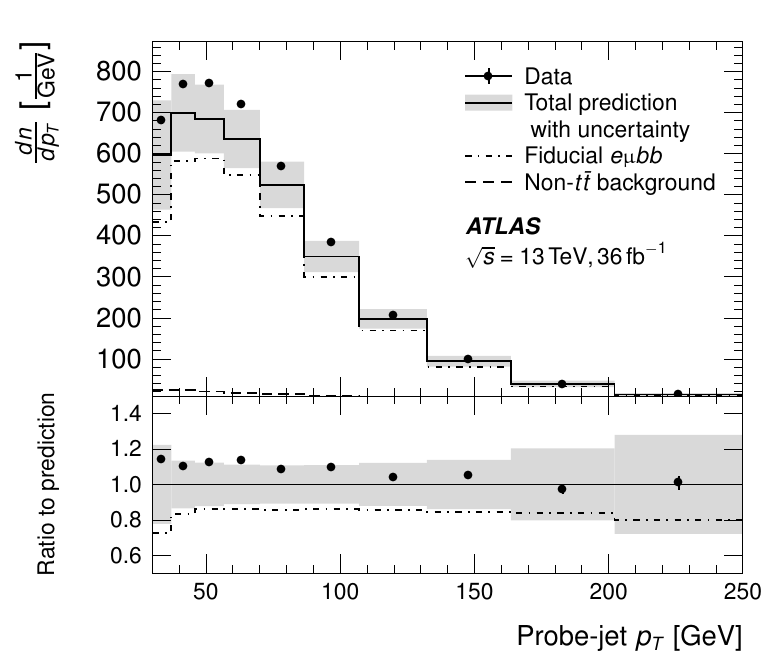}
}
 
\caption{
Comparison of detector-level probe-jet observable distributions between
simulation and collision data: (a) probe-jet $\eta$ and (b) \pt.
The nominal non-\ttbar background and fiducial $\ttbar \to e\mu bb$
predictions are shown in addition to the total prediction; the fiducial
probe-jet histogram is not stacked on top of the non-\ttbar background, in
order to show the expected fraction of fiducial events.
All systematic uncertainties are included in the uncertainty on the total prediction.
The first and last histogram bins do not include the underflow and overflow entries.
} \label{fig:probejetkinematics}
\end{figure}

\begin{figure}[h!]
\centering
 
\subfigure[][]{
\includegraphics[width=0.45\textwidth]{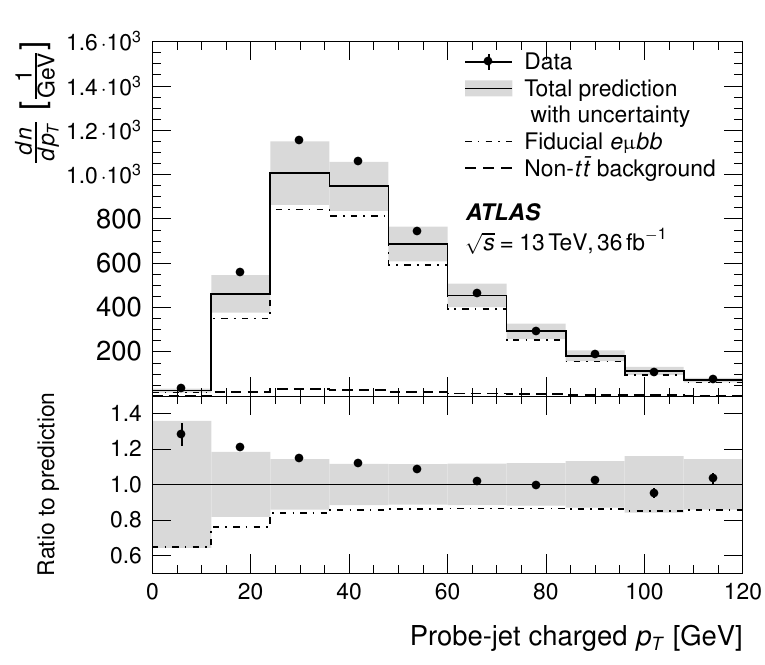}
}
\subfigure[][]{
\includegraphics[width=0.45\textwidth]{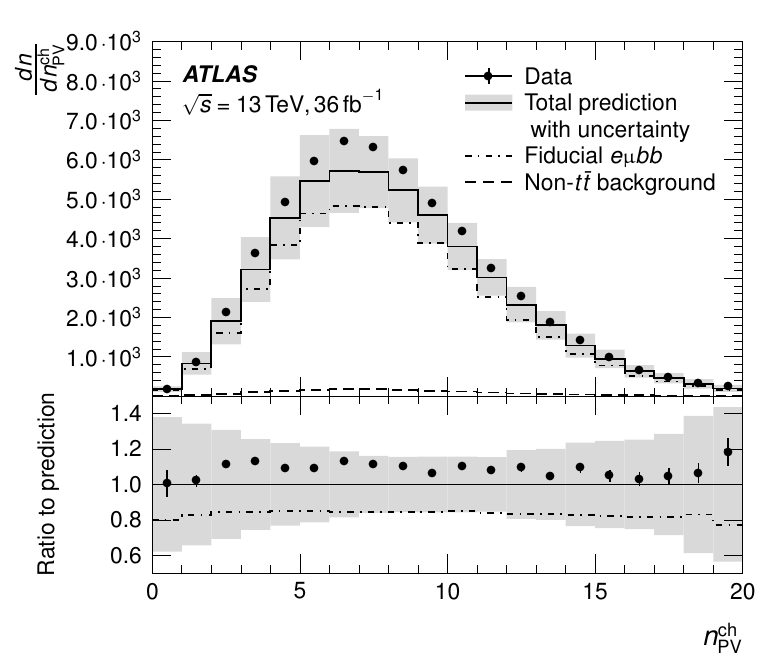}
}
\caption{
Comparison of detector-level probe-jet observable distributions between
simulation and collision data: (a) the probe-jet total charged \pt and (b) the
number of tracks matched to the jet that originate from the primary vertex.
The nominal non-\ttbar background and fiducial $\ttbar \to e\mu bb$
predictions are shown in addition to the total prediction; the fiducial
probe-jet histogram is not stacked on top of the non-\ttbar background, in
order to show the expected fraction of fiducial events.
All systematic uncertainties are included in the uncertainty on the total prediction.
The first and last histogram bins do not include the underflow and overflow entries.
} \label{fig:probentrk}
\end{figure}

\begin{figure}[h!]
\centering
 
\subfigure[][]{
\includegraphics[width=0.45\textwidth]{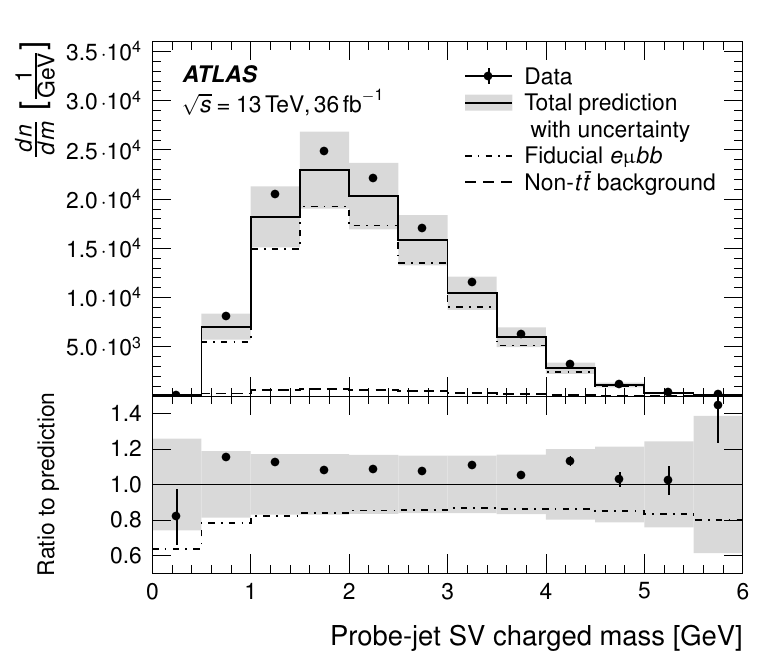}
}
\subfigure[][]{
\includegraphics[width=0.45\textwidth]{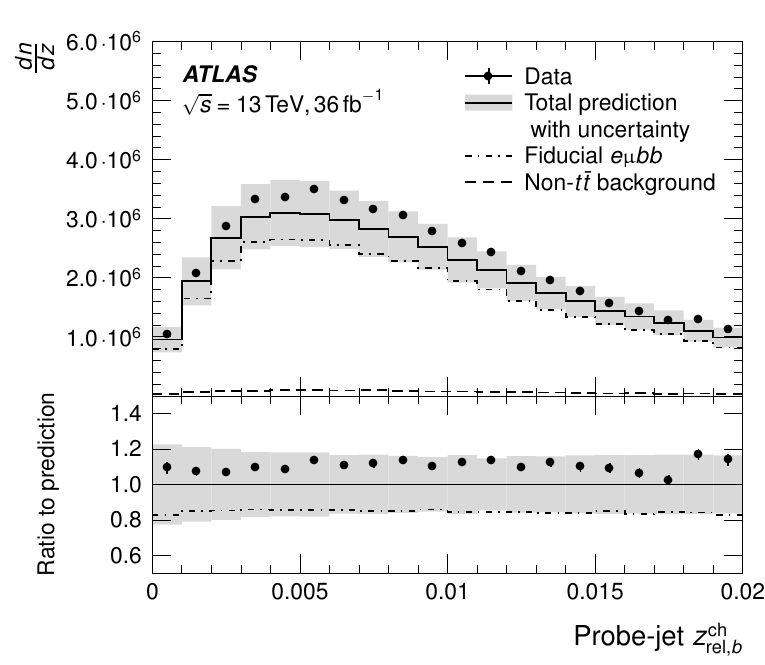}
}
 
\caption{
Comparison of detector-level probe-jet observable distributions between
simulation and collision data: (a) the invariant mass of all track constituents of
the jet secondary vertex and (b) the relative momentum of the secondary vertex
transverse to the jet charged momentum, $\zbrelc = |(\pbcvec \times \pjcvec)|
/ |\pjcvec|^2$.
The nominal non-\ttbar background and fiducial $\ttbar \to e\mu bb$
predictions are shown in addition to the total prediction; the fiducial
probe-jet histogram is not stacked on top of the non-\ttbar background, in
order to show the expected fraction of fiducial events.
All systematic uncertainties are included in the uncertainty on the total prediction.
The first and last histogram bins do not include the underflow and overflow entries.
} \label{fig:probesvmass}
\end{figure}

Table~\ref{tab:yields} shows the numbers of selected events and probe-jets after
the requirements from Section~\ref{ssec:selection} in both the MC prediction and
the observed collision data, including all uncertainties described in
Section~\ref{sec:systematics}.
Figures~\ref{fig:probejetkinematics}--\ref{fig:probesvmass} compare relevant
probe-jet observables between the MC prediction and the observed data. In both
cases, all uncertainties described in Section~\ref{sec:systematics} are included
in the prediction uncertainty.
Agreement within the estimated uncertainties is observed between the simulated
and observed events, showing that effects from the underlying event, pileup,
track reconstruction, and secondary-vertex finding are well-modeled by the
simulation for the purpose of $b$-hadron and $b$-jet measurements.

\begin{table}
\centering
\caption{
Summary of expected and observed event and probe-jet yields.
The uncertainties on the predicted yields include all components discussed in
Section~\ref{sec:systematics}.
Note that the sums of column entries may not agree exactly with reported total
yields due to rounding.
}
\label{tab:yields}
\vspace{0.2cm}
\begin{tabular}{ l | c | c }
\toprule
& Events with $e\mu jj$ ($\ge 1$ $b$-tag)  & Probe-jets  \\
\hline
Process                 & \multicolumn{2}{c}{Predicted yields} \\
\hline
Fiducial \ttbar         & ~~--                &   $44000 \pm 9000$  \\
Nonfiducial \ttbar      & ~~--                &  ~~$6700 \pm 1500$   \\
\hline
Total \ttbar            & ~~$76000 \pm 12000$ &  $51000 \pm 9000$  \\
\hline
Single top              & ~~$4400 \pm 1500$   &  $1580 \pm 600$    \\
$Z$+jets                & $125 \pm 45$        &  $13.0 \pm 5.1$    \\
Diboson                 & ~~$90 \pm 34$       &  ~~$9.7 \pm 3.9$    \\
\hline
Total non-\ttbar\       & ~~$4600 \pm 1600$   &  $1600  \pm 600$    \\
\hline
$b$-jets                & ~~--                &  $52000 \pm 9000$  \\
$c$-jets                & ~~--                &  $180 \pm 60$     \\
Other jets              & ~~--                &  $250 \pm 70$     \\
\hline
Total prediction        & ~~$81000 \pm 13000$  &  $53000 \pm 9000$  \\
\hline
& \multicolumn{2}{c}{Observed yields} \\
\hline
Data                     & 88511~~~~~~~~~~~~\,          &  57476~~~~~~~~~~~~\,              \\
\bottomrule
\end{tabular}
\end{table}

\FloatBarrier
% End of text imported from the .//analysis.tex input file
 
% The next lines are included from the .//unfolding.tex input file
\section{Unfolding procedure}
\label{sec:unfolding}
 
The FBU technique is used to extract the posterior probability of the
particle-level differential cross-sections given the observed
data~\cite{Choudalakis:2012hz}.
A likelihood of the data given a prediction is constructed with the
particle-level signal cross-sections (\xspl) as the parameters of interest.
Systematic uncertainties, which are described in more detail in
Section~\ref{sec:systematics}, are included in the likelihood as nuisance
parameters (NPs), whose vector of values is denoted by \nps.
 
To build the likelihood function, the predicted numbers of signal
and background events is calculated in each bin of the detector-level
observables (defined in Section~\ref{ssec:observables}), given a set of \xspl\
and \nps\ values.
The total predicted event count, $x_i$, in bin $i$ of a detector-level
observable is first constructed as the luminosity, $\lumi(\nps)$, times the
predicted background cross-section, $b_i(\nps)$, plus the response matrix times
the particle-level signal cross-sections,
$\sum_j \respm_{ij}(\nps) \sigma^\mathrm{p}_j$, where $j$ runs over the
particle-level bin indices,
 
\begin{equation*}
x_i(\xspl, \nps) = \lumi(\nps) \cdot
\Big(b_i(\nps) + \sum_j \respm_{ij}(\nps) \sigma_j^\mathrm{p}\Big),
\end{equation*}
where the response matrices $\respm_{ij}(\nps)$, background predictions
$b_i(\nps)$, and their related uncertainties are derived from simulation.
 
The background cross-section predictions $b_i(\nps)$ are defined as
 
\begin{equation*}
b_i(\nps) = b_{i,0} + \sum_{k \in \, \mathrm{NPs}}
\lambda_k (b_{i,k} - b_{i,0}) \, ,
\end{equation*}
 
where $b_{i,0}$ predicts the nominal background in bin $i$, and $b_{i,k}$
predicts the background in bin $i$ with the $1\sigma$ variation corresponding to
nuisance parameter $k$.
Detector response matrices are computed similarly:
 
\begin{equation*}
\respm_{ij}(\nps) = \respm_{ij,0} + \sum_{k \in \, \mathrm{NPs}}
\lambda_k (\respm_{ij,k} - \respm_{ij,0}) \, ,
\end{equation*}
 
where $\respm_{ij,0}$ is the nominal probability of a jet in particle-level bin
$j$ to be observed in detector-level bin $i$, and $\respm_{ij,k}$
is the corresponding probability for systematic variation $k$.
A likelihood is then constructed as the product of Poisson probabilites over
all detector-level bins $i$ as a function of the model parameters, \xspl and
\nps:
 
\begin{equation*}
\llh(\vec{d} | \xspl, \nps) = \prod_i
\mathrm{Poisson}(d_i, x_i(\xspl, \nps)).
\end{equation*}
 
Finally, Gaussian priors with $\mu = 0$ and $\sigma = 1$ are imposed for the
nuisance parameters corresponding to systematic variations; the one exception is
the luminosity, for which a log-normal prior is used with $\mu = 0$ and
$\sigma = 0.021$, corresponding to the luminosity uncertainty described in
Section~\ref{sec:systematics}.
A flat, non-negative prior is imposed on signal cross-sections, \xspl.
The posterior probability distribution given the observed data is then
 
\begin{equation*}
P(\xspl, \nps | \vec{d}) \propto \llh(\vec{d} | \xspl, \nps) \cdot
\prod_{k \in \mathrm{NPs}} \mathrm{Prior}_k (p_k).
\end{equation*}
 
The maximum of this distribution over all parameters of interest and nuisance
parameters is determined using gradient ascent.
The distribution is then marginalized by integrating over the nuisance
parameters via sampling with a Hamiltonian Markov chain Monte
Carlo~\cite{DUANE1987216}.
 
The dominant causes of imperfect detector response are tracking
reconstruction inefficiencies, $b$-hadron daughter tracks not being
correctly matched to the reconstructed secondary vertex, and tracks
from pileup particles being incorrectly matched to the hard-scatter
vertex.
This results in a shift in the observables' mean value of about $-0.2$ for \zblc
and \zbtc, and about $-1$ for \nsvtrk, which are accounted for during unfolding.
 
The unfolding procedure was found to correctly reproduce particle-level
spectra from detector-level observations.
To check this, a ``stress test'' of the unfolding was performed: detector-level
pseudodata from \HERWIG[7.0.4], \SHERPA[2.1.1], and \POWPY[8] with the A14 tune
\alphas variations were unfolded to the particle level.
In all cases the central values of the particle-level posterior distributions
were in excellent agreement with the true particle-level spectra, well within
the total uncertainty of the unfolded pseudodata.
% End of text imported from the .//unfolding.tex input file
 
% The next lines are included from the .//systematics.tex input file
\section{Systematic uncertainties}
\label{sec:systematics}
 
Systematic uncertainties associated with the detector, reconstruction, and simulation
are treated in this analysis by calculating the impact of each uncertainty on
the background prediction and detector response matrix and including these
variations in the unfolding model laid out in Section~\ref{sec:unfolding}.
Here, the sources of systematic uncertainty considered in the unfolding of
the measured observables are outlined.

\subsection{Theory and modeling uncertainties}
 
Six variations of the nominal \ttbar model are considered. Their
effects on both the response matrix and the predicted
nonfiducial \ttbar\ background are taken into account.
In addition to the sources of uncertainty listed below, an uncertainty in
the \ttbar normalization in this phase space is considered.
Recent experimental measurements in the exclusive $\ttbar + 0$-jet phase
space~\cite{Aaboud:2018uzf} quote uncertainties at the 10\% level, so
a 10\% uncertainty is assigned to the $\ttbar + 0$-jet cross-section.
Uncertainties from the \ttbar\ matrix-element calculation and matching scheme
were found to be small, as expected, and are neglected in the final analysis.
 
\begin{itemize}
\item Parton shower and hadronization:
\POWHER[7.0.4] is used to build an
alternative prediction and its deviation from the nominal
prediction is taken as an uncertainty.
\item Initial-state radiation:
\textsc{Var3} of the A14 \PYTHIA[8] tune corresponds to variations of $\alphas(m_Z)$
between 0.115 and 0.140. This parameter is varied together with the hadronization
and factorization scales in the matrix element and the $h_{\textrm{damp}}$ value as
described in Section~\ref{sec:mcsamples}, to increase and decrease
the amount of initial-state radiation in \ttbar events.
\item Final-state radiation:
The nominal \POWPY[8] A14 sample is compared with two samples generated with \POWPY[8] using the \textsc{Var2}
eigentune variations. The latter include the effects of varying several parameters in \PYTHIA[8], the largest
of which is due to varying of the amount of final-state radiation by shifting  $\alpha_\text{s}^{\text{FSR}}$ to 0.111 and 0.139, respectively.
\item Production and decay fractions of $b$-hadrons in \ttbar events:
To assess uncertainties
in the $b$-hadron species production fractions, the procedure
developed in a recent ATLAS measurement of the top-quark
mass~\cite{ATLAS-CONF-2019-046} is followed.
The $b$-hadron species production fractions and decay fractions in nominal
simulation are reweighted to experimental world averages~\cite{Zyla:2020zbs},
and effect on the detector response matrix is taken as an uncertainty.
\item Modeling of SV track multiplicity:
The secondary-vertex track multiplicity distribution in probe jets is compared
between the nominal prediction and data, as shown in Figure~\ref{fig:recoobs}
(d).
Weights are derived from any observed differences and applied to the
simulation to achieve agreement with the data.
The effect of this reweighting is propagated into the response matrix
as a systematic uncertainty of the detector response.
\item Top-quark and \bjet\ \pt\ modeling:
The jet charged \pt\ spectrum for the nominal prediction is compared with the
observed data, as shown in Figure~\ref{fig:probentrk} (a).
Weights are derived from the observed differences, and the effect of this
reweighting is propagated into the response matrix and treated as a systematic
variation.
\end{itemize}
 
\subsection{Non-\ttbar modeling uncertainties}
In addition to uncertainties in the prediction of \ttbar events, a 30\% uncertainty
is assessed for the normalization of three non-\ttbar background processes
with two prompt leptons: $tW$, Drell--Yan, and diboson production.
The largest of these backgrounds is from $tW$ production, for which  the difference between the
diagram removal (DR) and diagram subtraction (DS) schemes for calculating
interference between single-top-quark and \ttbar production is also
considered~\cite{Frixione:2008yi}.

\subsection{Detector uncertainties}
\label{sssec:detectoruncerts}
 
The systematic uncertainties related to imperfect understanding of the
detector in data have an impact on the estimated background yield and on the
signal detector response.
The jet energy scale uncertainty consists of 18~eigenvector components and
is derived with a combination of test-beam data, in situ
measurements and simulation at 13~\TeV~\cite{Aaboud:2017jcu}.  Further contributions
originate from the $\eta$-intercalibration, jet-flavor composition and
response, single-particle response and pileup effects. An in situ
measurement of the jet response in dijet events~\cite{JETM-2018-05} was used to
estimate the systematic uncertainty due to the jet energy resolution.
An additional uncertainty from the efficiency of the JVT in removing pileup jets is evaluated.
Since measured observables are built from charged-particle tracks, only the overall
event reconstruction efficiency is significantly affected by calorimeter jet uncertainties.
 
The uncertainties related to lepton identification, reconstruction, isolation
and trigger efficiencies are taken into account by a variation of the
corresponding scale factors in simulation within their assigned uncertainties.
Additional uncertainties for the modeling of the lepton energy/momentum
resolution and scale, obtained from measurements in data using $J/\psi \to \ell
\ell$ and $Z \to \ell \ell$ events~\cite{PERF-2017-01,Aad:2016jkr} are also
considered.
Their impacts on the predicted background yields and detector response are taken
into account.
 
To correct for different $b$-tagging efficiencies in simulation and data, scale
factors are applied to the simulation. Different scale factors have been
obtained for $b$-jets, $c$-jets, and light-parton jets from \ttbar data, dijet
data, and simulated events~\cite{Aad:2015ydr,ATLAS-CONF-2018-001,ATLAS-CONF-2018-006}, which are then combined to give
nominal scale factors and associated uncertainties. The total uncertainty due to the
$b$-tagging efficiency and fake rate in this analysis was obtained by
varying these scale factors within their respective uncertainties.
 
Uncertainties due to differences between the simulation and data in the detector
alignment, track reconstruction efficiency, track fake rate, and track
impact parameter ($d_0$ and $z_0$) resolutions are determined using
$Z\rightarrow \mu\mu$, dijet, and minimum-bias events~\cite{PERF-2015-08,ATL-PHYS-PUB-2015-051}.
The distribution of the average number of interactions per bunch crossing
($\langle \mu \rangle$) is altered to assess the effect of possible mismodeling of additional pileup
vertices on the measurement~\cite{LUCID2,ATLAS-CONF-2019-021}.
The uncertainty in the combined 2015--2016 integrated luminosity is
2.1\%~\cite{ATLAS-CONF-2019-021}, obtained using the LUCID-2 detector~\cite{LUCID2}
for the primary luminosity measurements.
 
\subsection{Uncertainty pruning}
\label{sssec:uncertpruning}
 
Systematic uncertainties with a negligible impact on the background prediction
and response matrices are not included in the unfolding procedure;
variations that change the predicted background by less than 5\% in all
detector-level observable bins and alter the response matrix by less than 0.01\%
in all bins are removed.
Tracking uncertainties from the alignment of the inner detector, and uncertainties in
the lepton momentum scales and reconstruction efficiencies,
the JVT efficiency for hard-scatter jets, the charm- and light-flavor jet $b$-tagging mistag rates,
the normalization of the $Z+\mathrm{jets}$ and diboson production cross-sections,
and the predicted yield of events with only one prompt lepton are all
found to have only a small impact on the unfolding ingredients and are therefore
neglected.
% End of text imported from the .//systematics.tex input file
 
% The next lines are included from the .//results.tex input file
\section{Results}
\label{sec:results}
 
\begin{figure}[h!]
\centering
 
\subfigure[][]{
\includegraphics[width=0.45\textwidth]{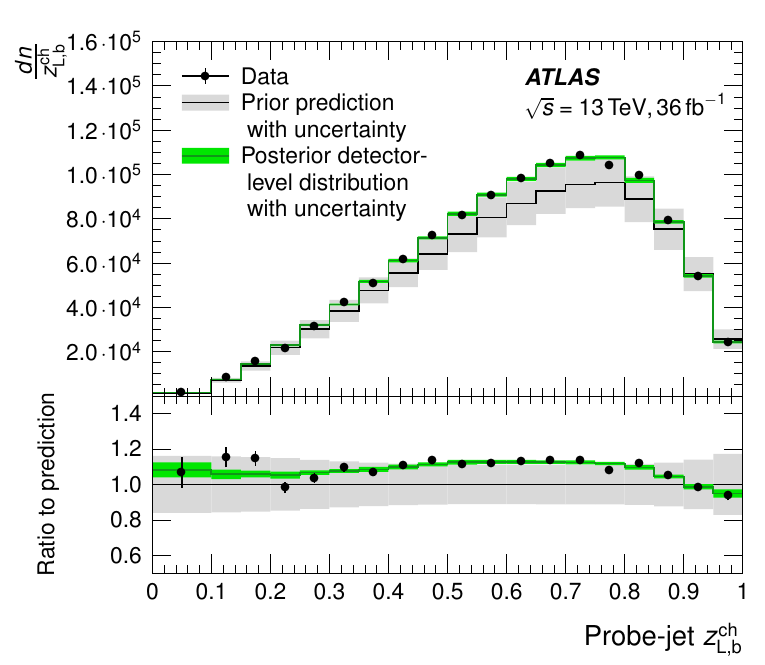}
}
\subfigure[][]{
\includegraphics[width=0.45\textwidth]{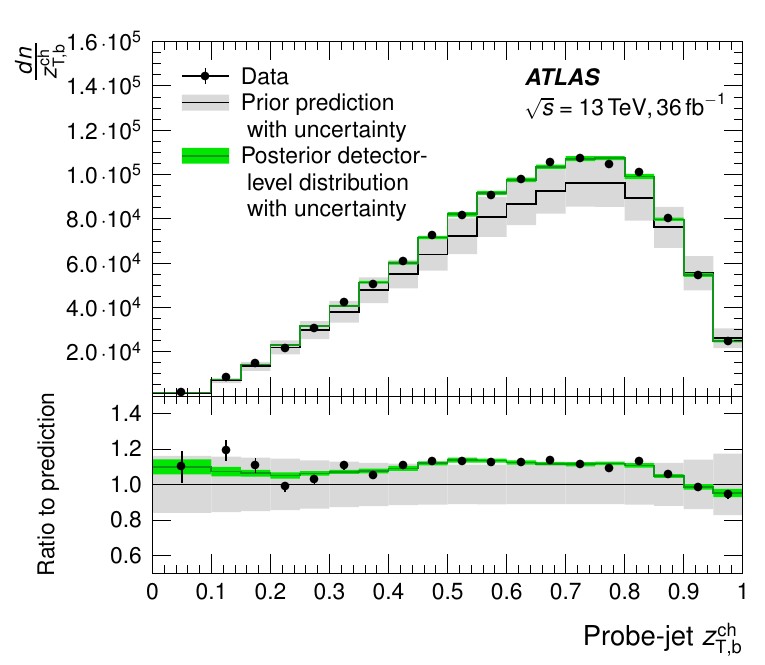}
} \\
\subfigure[][]{
\includegraphics[width=0.45\textwidth]{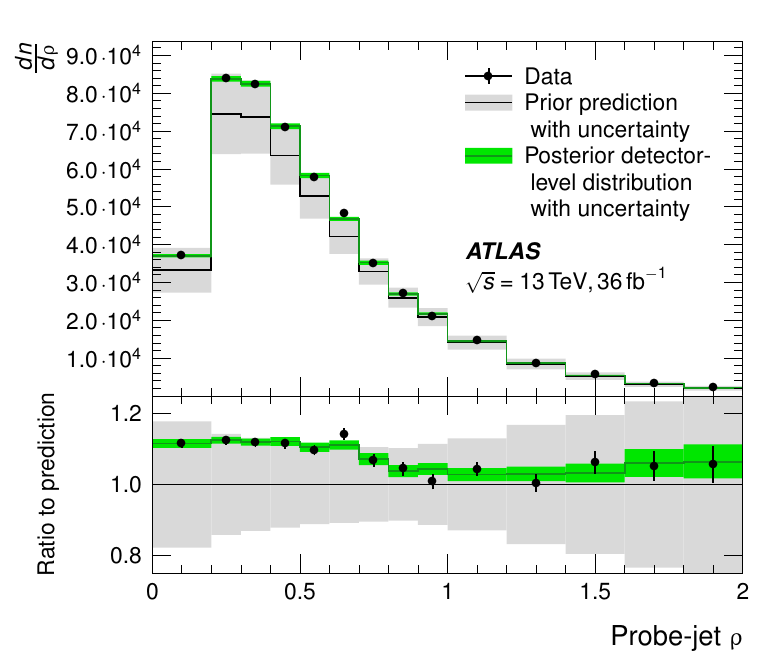}
}
\subfigure[][]{
\includegraphics[width=0.45\textwidth]{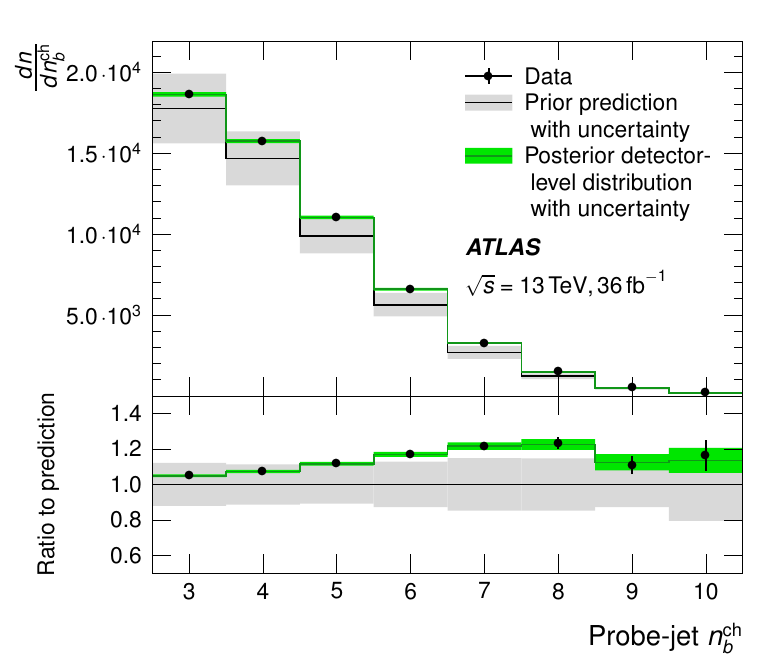}
}
 
\caption{
Comparison between simulation and collision data for detector-level
observables of interest: (a) \zblc, (b) \zbtc, (c) $\rho$, and (d) \nsvtrk.
The ``prior'' curve corresponds to the nominal observable prediction before
unfolding; only uncertainties due to background and detector modeling are
included in the uncertainty band.
The ``posterior'' curve corresponds to the posterior probability at detector-level
given the observed data used to unfold to particle level.
Good agreement is observed between the posterior distribution and the
observed data, and the posterior bands reflect the data statistical uncertainty,
indicating that the unfolding model is able to describe the detector-level
data.
} \label{fig:recoobs}
\end{figure}

Detector-level observable spectra for the collected data and the prediction, both
before and after unfolding, are shown in Figure~\ref{fig:recoobs}; all
uncertainties in the background prediction and signal detector response, as
described in Section~\ref{sec:systematics}, are included in the uncertainty bands.
Reasonable agreement between the detector-level data and simulation is observed
for \zblc, \zbtc, and $\rho$.
However, the prior prediction shows slightly fewer SV tracks than appear in the
data, which motivates the data-based $b$-hadron production and decay fraction
uncertainty described in Section~\ref{sec:systematics}.
Based on the level of agreement between the data and the posterior
detector-level distribution, it is concluded that the model used to unfold the
detector effects is sufficient to describe the observed data.
The large difference between the prior and posterior uncertainty bands for the
detector-level observables is driven by correlations between parameters in the
posterior model, which are introduced during the fit to data.

\begin{figure}[h!]
\centering
 
\subfigure[][]{
\includegraphics[width=0.45\textwidth]{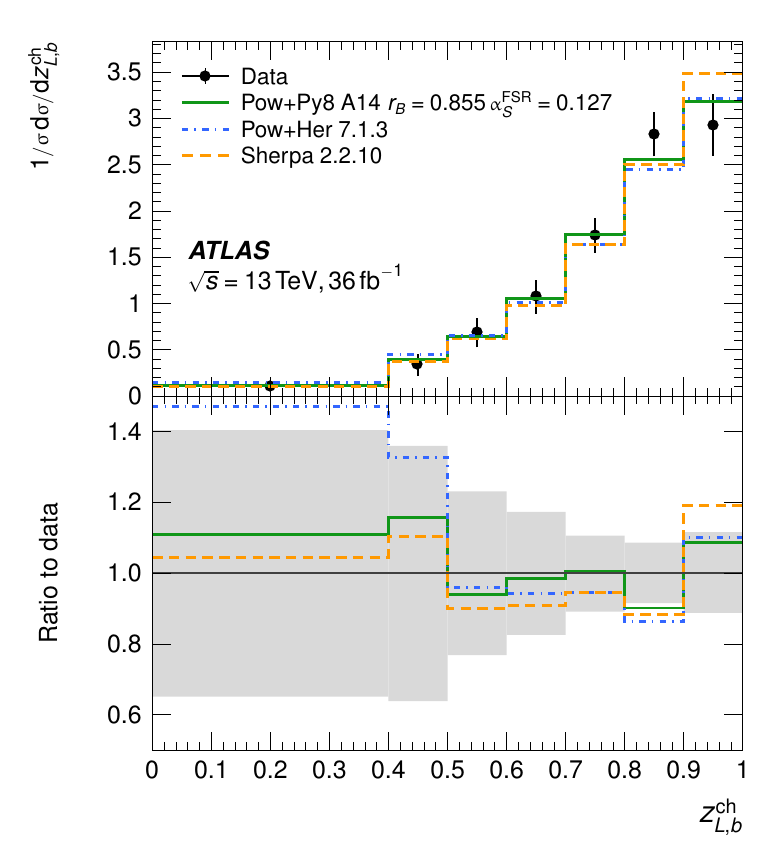}
}
\subfigure[][]{
\includegraphics[width=0.45\textwidth]{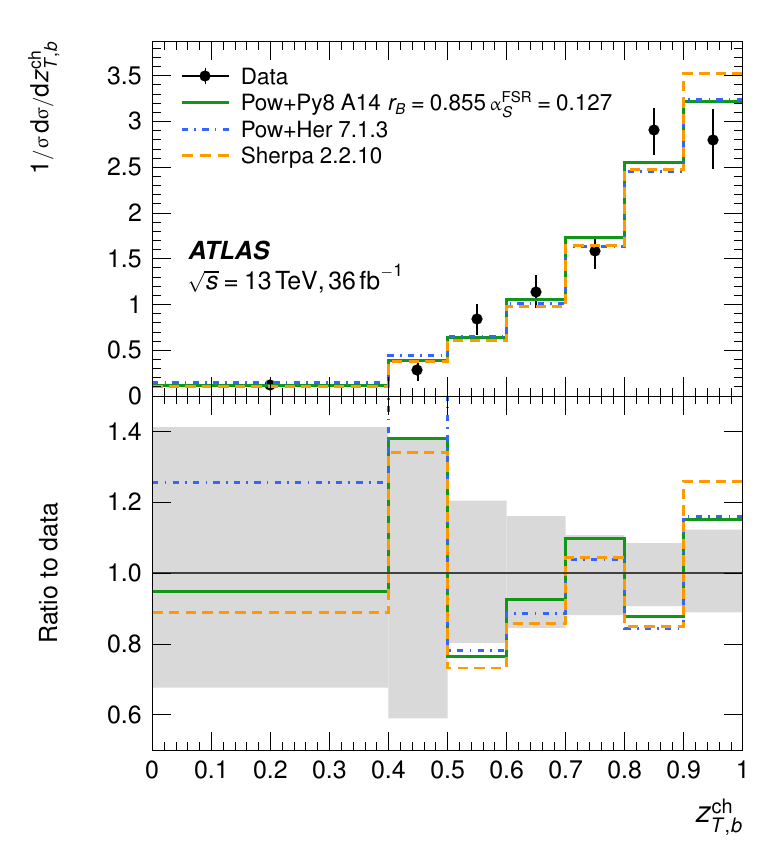}
} \\
\subfigure[][]{
\includegraphics[width=0.45\textwidth]{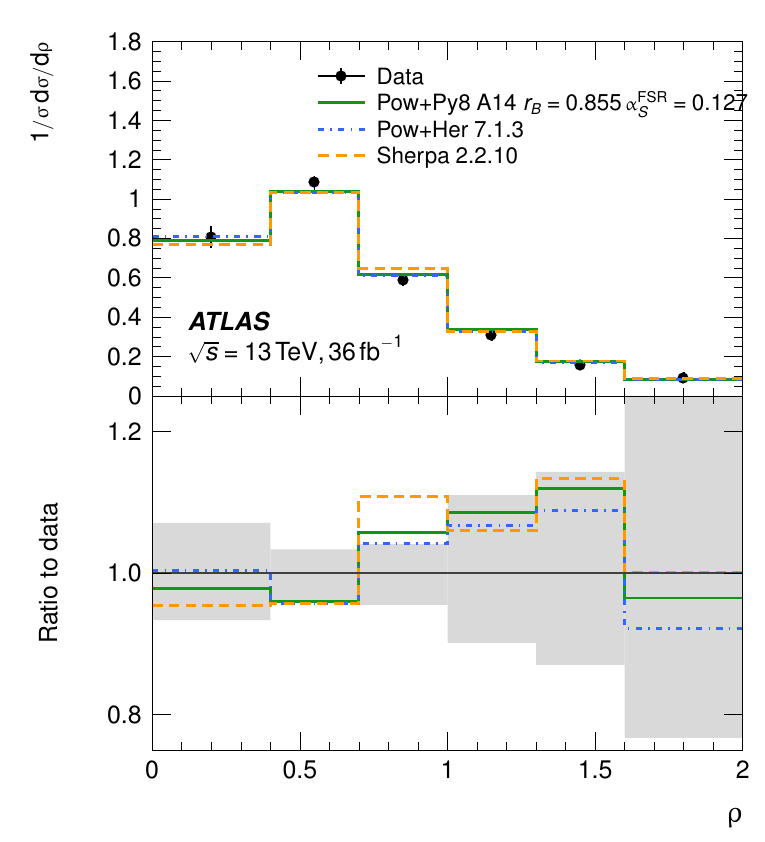}
}
\subfigure[][]{
\includegraphics[width=0.45\textwidth]{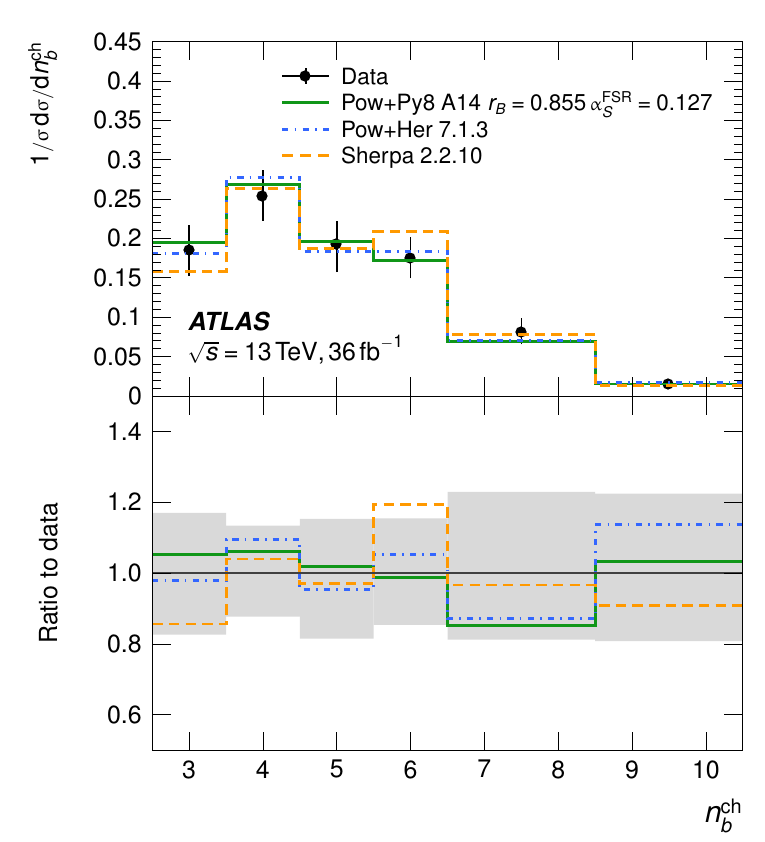}
}
 
\caption{
Comparison of particle-level observable distributions between MC and unfolded
data: (a) \zblc, (b) \zbtc, (c) $\rho$, and (d) \nsvtrk.
The plotted data points correspond to the maximum likelihood for the
particle-level cross-section, and the gray uncertainty band in the lower
panel of each figure shows the total uncertainty on the measured
cross-sections.
} \label{fig:plotsnormdata}
\end{figure}

\begin{figure}[h!]
\centering
 
\subfigure[][]{
\includegraphics[width=0.45\textwidth]{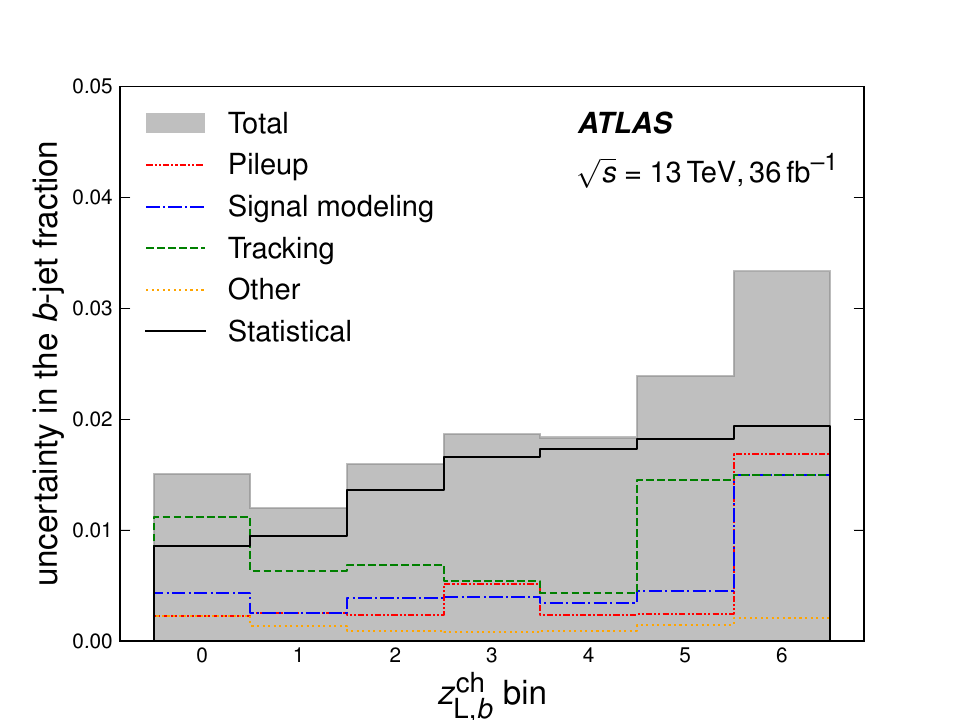}
}
\subfigure[][]{
\includegraphics[width=0.45\textwidth]{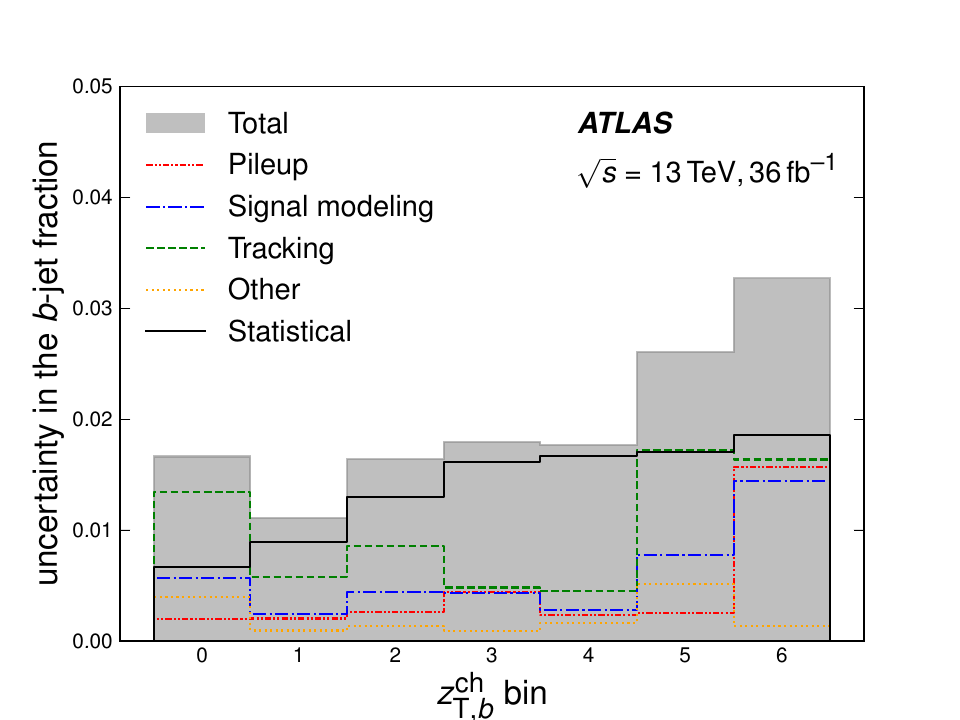}
} \\
\subfigure[][]{
\includegraphics[width=0.45\textwidth]{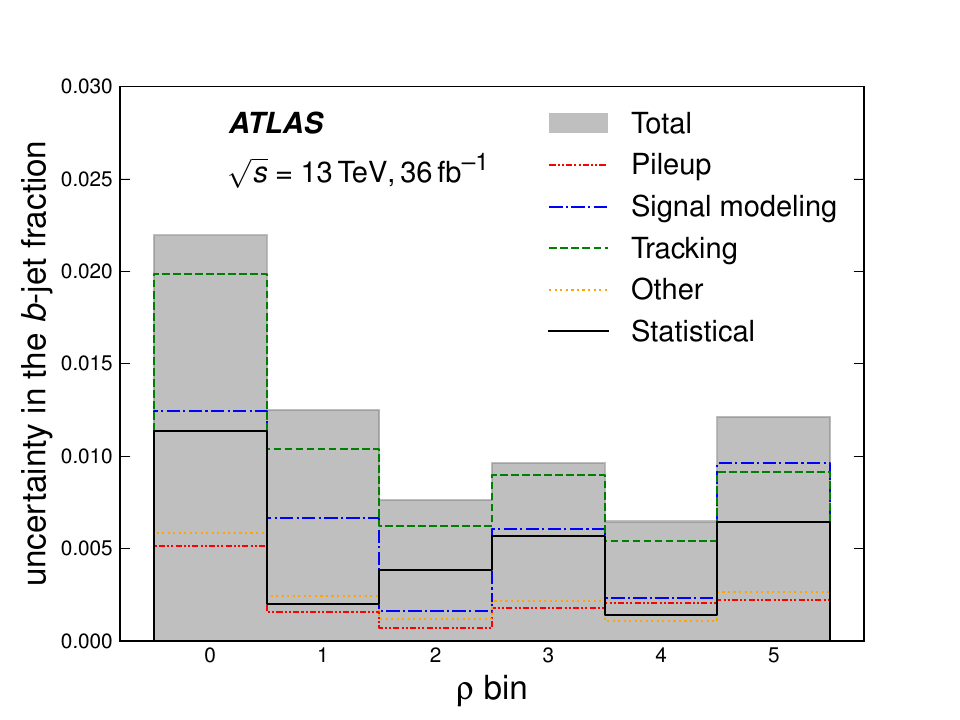}
}
\subfigure[][]{
\includegraphics[width=0.45\textwidth]{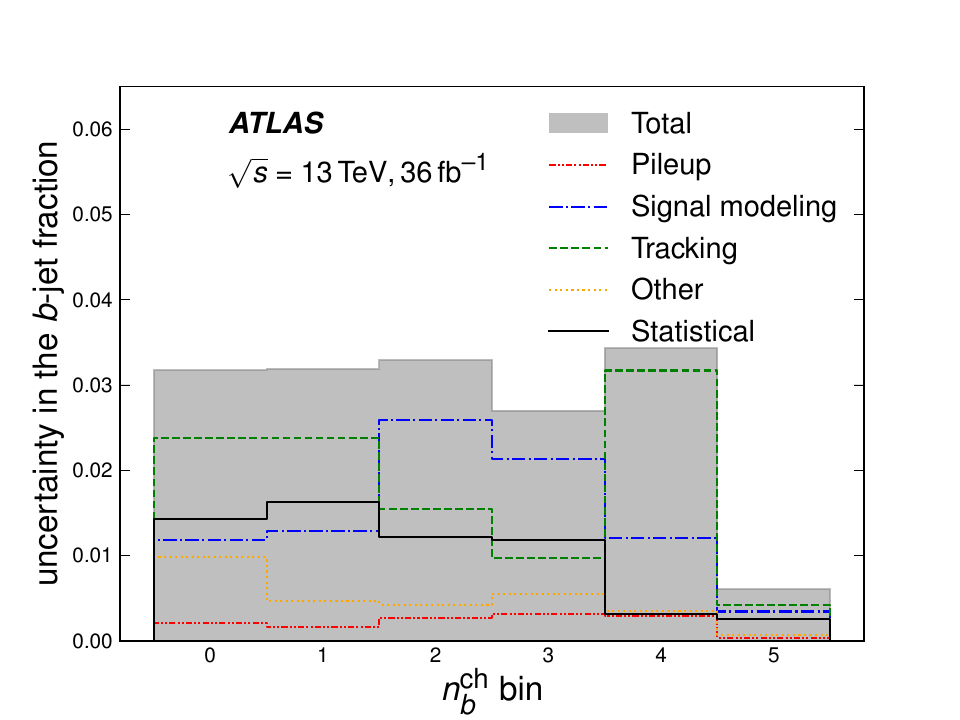}
}
\caption{
Sources of uncertainty in the $b$-jet fractions in each particle-level
observable bin for (a) \zblc, (b) \zbtc, (c) $\rho$, and (d) \nsvtrk.
} \label{fig:uncertbreakdown}
\end{figure}

Posterior distributions for the particle-level cross-sections are extracted
using the observed detector-level data, predicted detector response matrices,
and predicted detector-level backgrounds. In all figures, data points correspond
to the maximum-likelihood particle-level cross-sections, and the uncertainty
bands are the 16\%--84\% quantiles of the posterior distribution.
Figure~\ref{fig:plotsnormdata} compares the unfolded particle-level observable
spectra with MC predictions from \POWPY[8], \POWHER[7], and \SHERPA[2], which
are commonly used in other ATLAS data analyses.
The unfolded spectra are summarized in Table~\ref{tab:obs_values}.
Figure~\ref{fig:uncertbreakdown} shows the systematic uncertainties of the
fiducial $b$-jet fractions broken down by source; the statistical uncertainty is
dominant in the middle bins of the \zbl\ and \zbt distributions but subdominant
elsewhere.

\begin{table}[h!]
\scriptsize
\centering
\caption{Summary of observed \bjet\ spectra in particle-level bins for the
four observables.}
\label{tab:obs_values}
\vspace{0.2cm}
\renewcommand{\arraystretch}{1.5}
\begin{tabular}{ l | c | c | c | c | c | c | c }
\hline
$\zblc$ range & 0.0--0.4 & 0.4--0.5 & 0.5--0.6 & 0.6--0.7 & 0.7--0.8 & 0.8--0.9 & 0.9--1.0 \\
\hline
Fraction &	$0.041	\pm 0.015$	&	$0.034	\pm 0.012$	&	$0.068	\pm 0.016$	&	$0.107	\pm 0.019$	&	$0.174	\pm 0.018$	&	$0.283	\pm 0.024$	&	$0.293	\pm 0.033$	\\
 
\hline
\hline
$\zbtc$ range & 0.0--0.4 & 0.4--0.5 & 0.5--0.6 & 0.6--0.7 & 0.7--0.8 & 0.8--0.9 & 0.9--1.0 \\
\hline
Fraction &	$0.048	\pm 0.017$	&	$0.028	\pm 0.011$	&	$0.083	\pm 0.016$	&	$0.114	\pm 0.018$	&	$0.157	\pm 0.018$	&	$0.289	\pm 0.026$	&	$0.281	\pm 0.033$	\\

\hline
\hline
$\rho$ range & 0.0--0.4 & 0.4--0.7 & 0.7--1.0 & 1.0--1.3 & 1.3--1.6 & 1.6--2.0 & \\
\hline
Fraction	&	$0.323	\pm 0.022$	&	$0.323	\pm 0.012$	&	$0.175	\pm 0.008$	&	$0.093	\pm 0.010$	&	$0.047	\pm 0.006$	&	$0.038	\pm 0.012$	\\
\hline
 
\hline
\hline
\nsvtrk range & 3 & 4 & 5 & 6 & 7--8 & 9--10 & \\
\hline
Fraction 	&	$0.185	\pm 0.032$	&	$0.255	\pm 0.032$	&	$0.190	\pm 0.033$	&	$0.176	\pm 0.027$	&	$0.164	\pm 0.034$	&	$0.030	\pm 0.006$	\\
\end{tabular}
\end{table}
 
\begin{figure}[h!]
\centering
 
\subfigure[][]{
\includegraphics[width=0.45\textwidth]{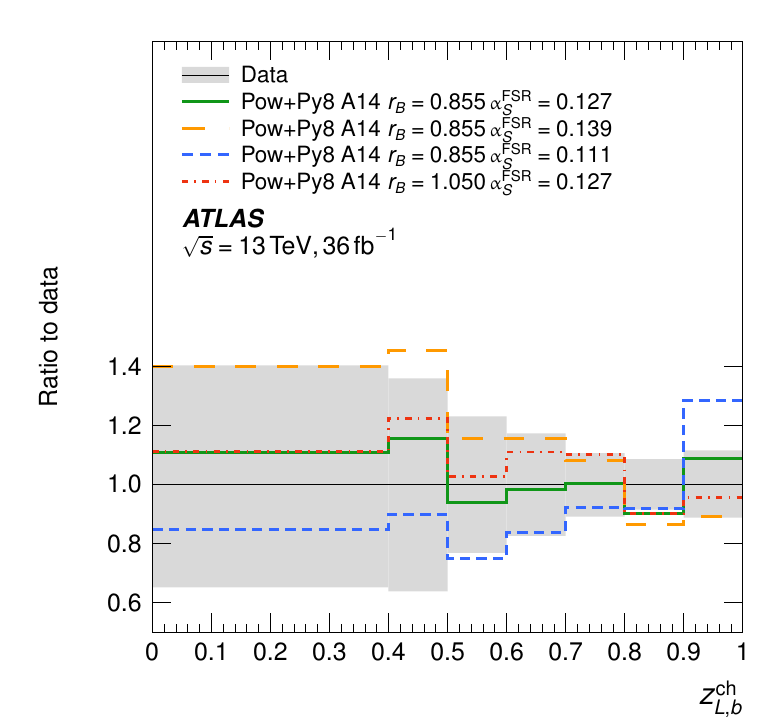}
}
\subfigure[][]{
\includegraphics[width=0.45\textwidth]{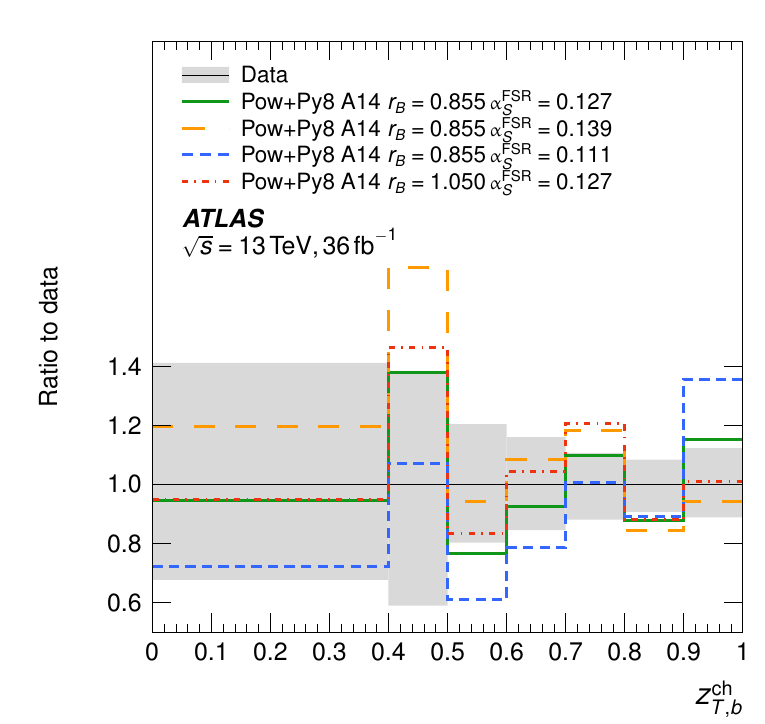}
} \\
\subfigure[][]{
\includegraphics[width=0.45\textwidth]{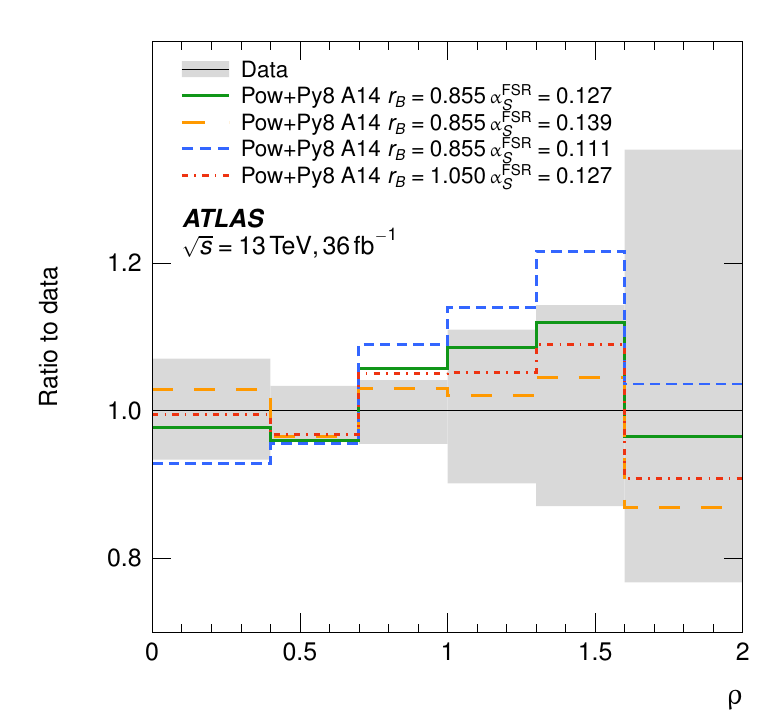}
}
\subfigure[][]{
\includegraphics[width=0.45\textwidth]{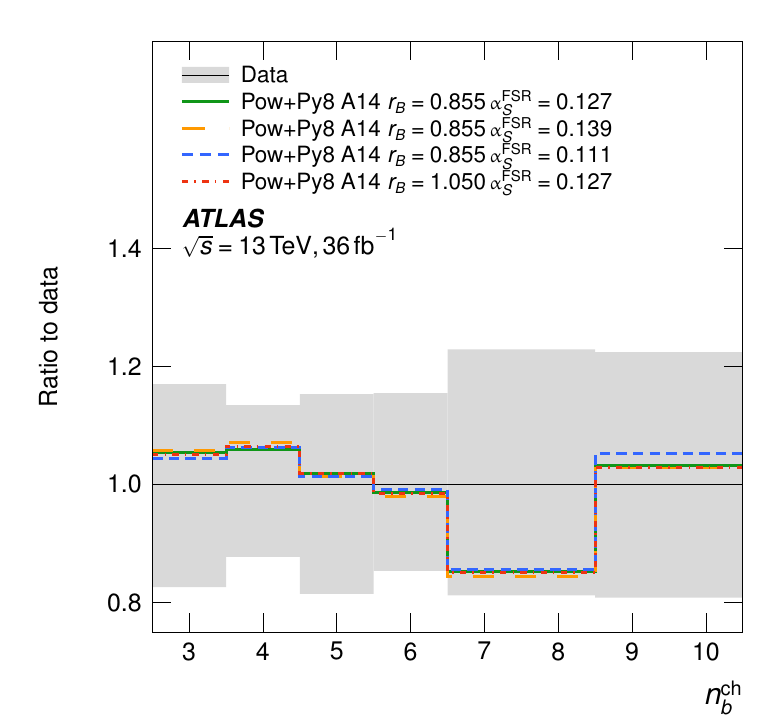}
}
\caption{Comparison of particle-level observables between \POWPY[8] A14
variations and unfolded data for (a) \zblc, (b) \zbtc, (c) $\rho$, and (d)
\nsvtrk.
} \label{fig:pythiaA14}
\end{figure}

\begin{figure}[h!]
\centering
 
\subfigure[][]{
\includegraphics[width=0.45\textwidth]{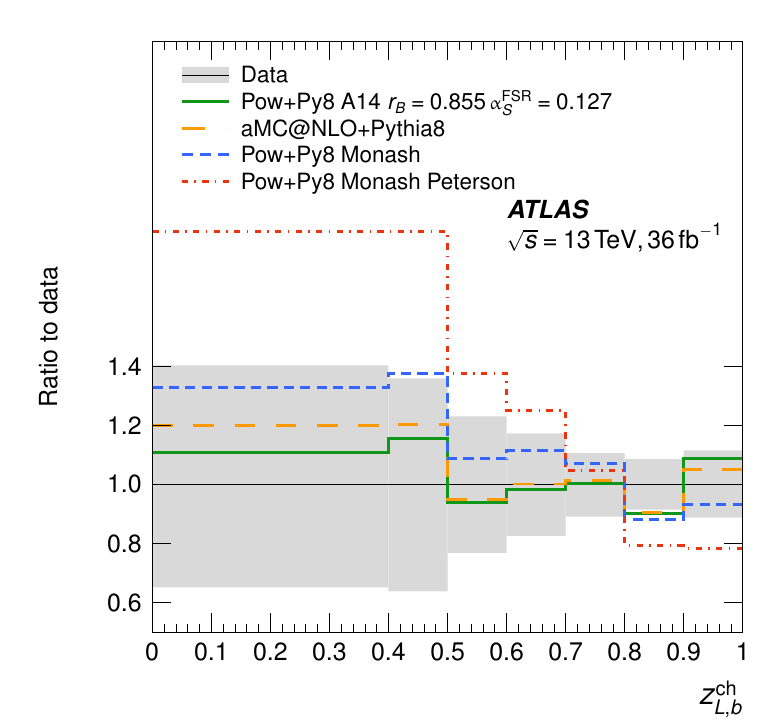}
}
\subfigure[][]{
\includegraphics[width=0.45\textwidth]{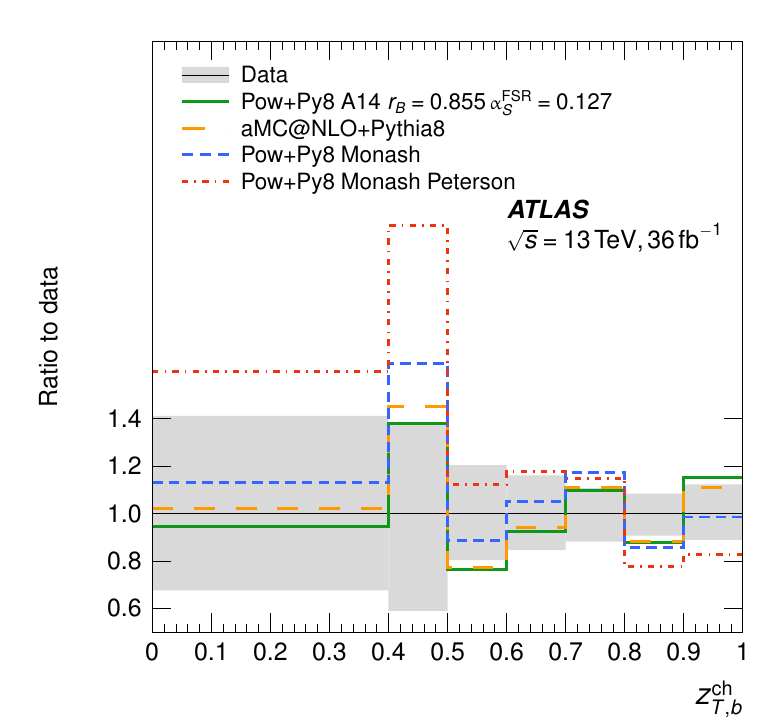}
} \\
\subfigure[][]{
\includegraphics[width=0.45\textwidth]{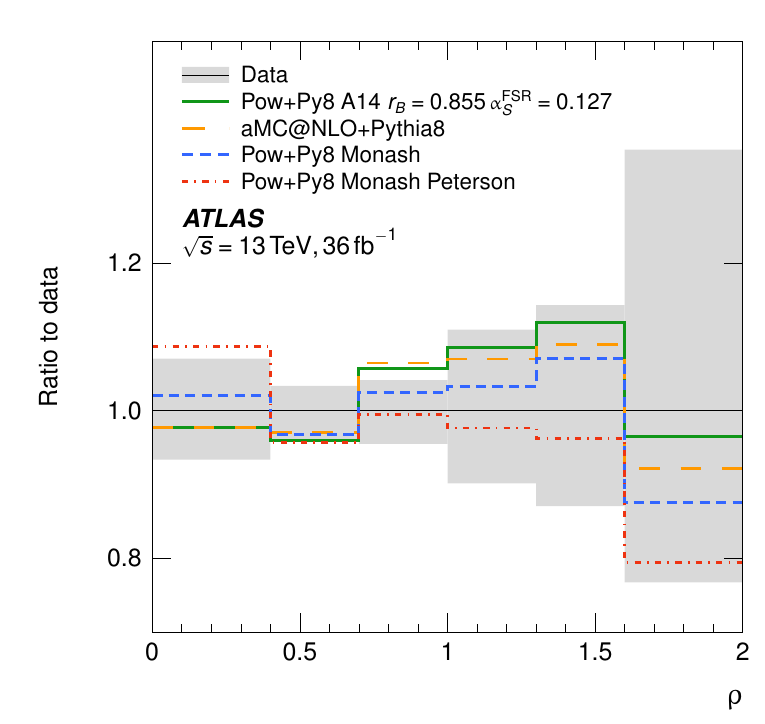}
}
\subfigure[][]{
\includegraphics[width=0.45\textwidth]{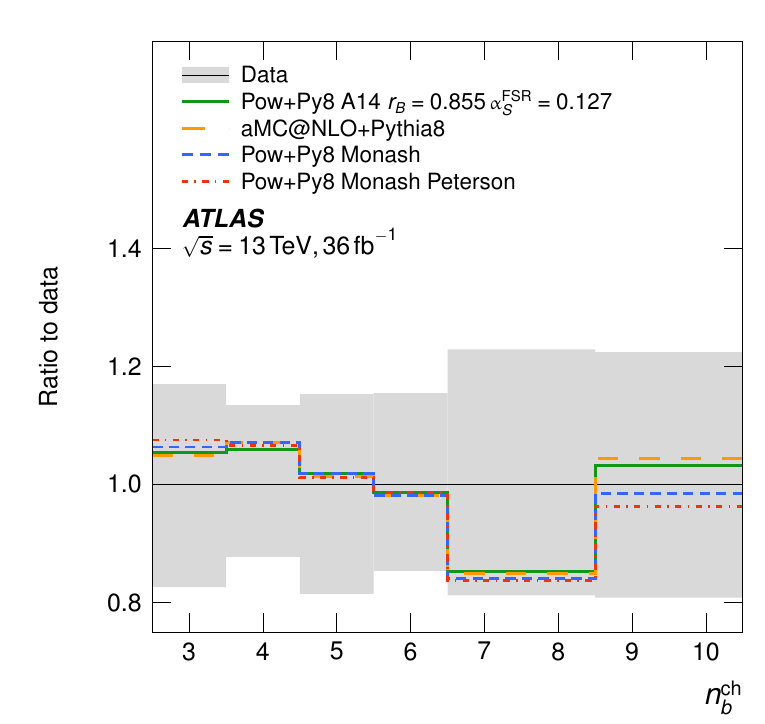}
}
\caption{Comparison of particle-level observables between \PYTHIA[8] tunes
and unfolded data for (a) \zblc, (b) \zbtc, (c) $\rho$, and (d) \nsvtrk.
} \label{fig:pythia}
\end{figure}

The effect on these observables of changing the values of nonperturbative MC parameters, which were
largely tuned to the LEP and SLC $e^+e^- \to b\bar{b}$ data, is explored.
Figure~\ref{fig:pythiaA14} compares the data with predictions using various choices of
\asfsr and $r_B$, a parameter rescaling the $b$-quark mass, in
the \PYTHIA[8] parton shower with the Lund--Bowler fragmentation
model~\cite{Bowler:1981sb}.
The \asfsr values 0.127, 0.111, and 0.139 are those used in the
nominal and down/up variations of the A14 tune, respectively; in the original
A14 tune, $r_B$ is taken from the Monash tune and set to
0.855~\cite{Skands:2014pea}.
In the context of measuring the top-quark mass, $r_B$ was recently
re-fitted to the electron--positron collider data with the nominal $\asfsr =
0.127$ value from the A14 tune; a value of $r_B = 1.050$ was found to be
optimal, and this choice is also compared with data in
Figure~\ref{fig:pythiaA14}~\cite{ATLAS-CONF-2019-046}.
For \zblc, the A14 \PYTHIA[8] tune and its \asfsr variations accurately
describe the data and provide a reasonable uncertainty envelope for $b$-quark
fragmentation.
Figure~\ref{fig:pythia} shows alternative \PYTHIA[8] setups that are in common
use: \amcpy, the Monash tune of \PYTHIA interfaced to \powheg, and the Monash
tune together with the Peterson fragmentation model~\cite{PhysRevD.27.105} again
interfaced to \powheg.
The Peterson model is found to be in poor agreement with the unfolded data for
the $z$ and $\rho$ observables.

\begin{figure}[h!]
\centering

\subfigure[][]{
\includegraphics[width=0.45\textwidth]{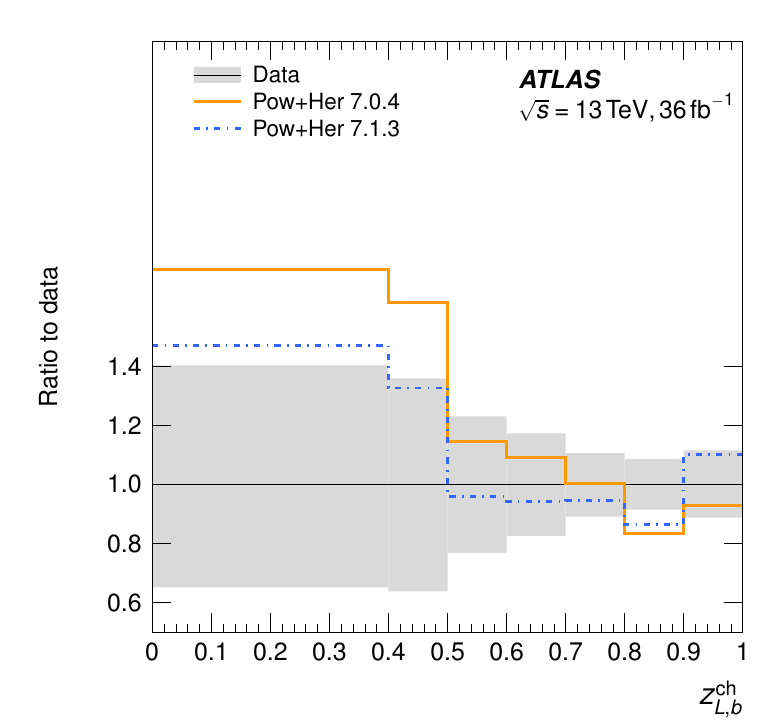}
}
\subfigure[][]{
\includegraphics[width=0.45\textwidth]{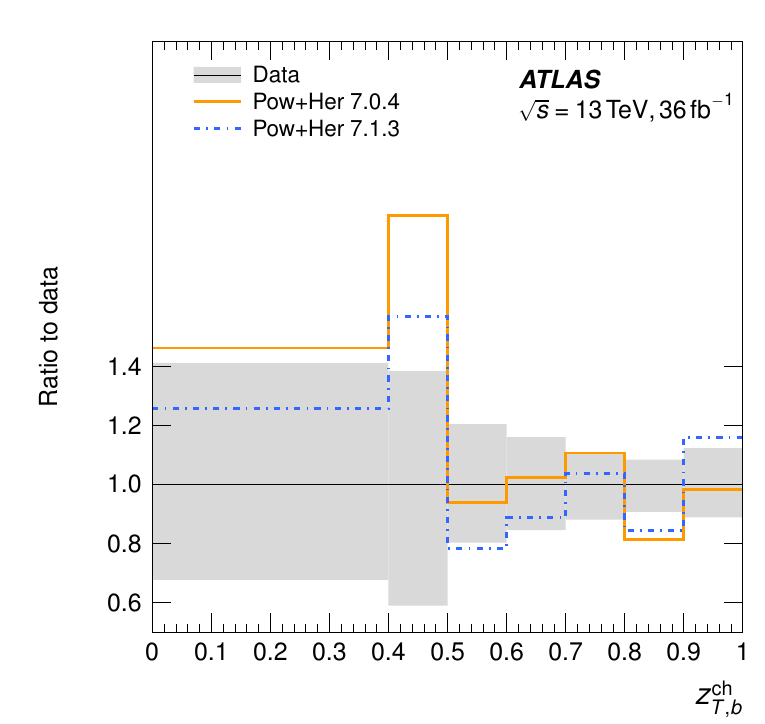}
} \\
\subfigure[][]{
\includegraphics[width=0.45\textwidth]{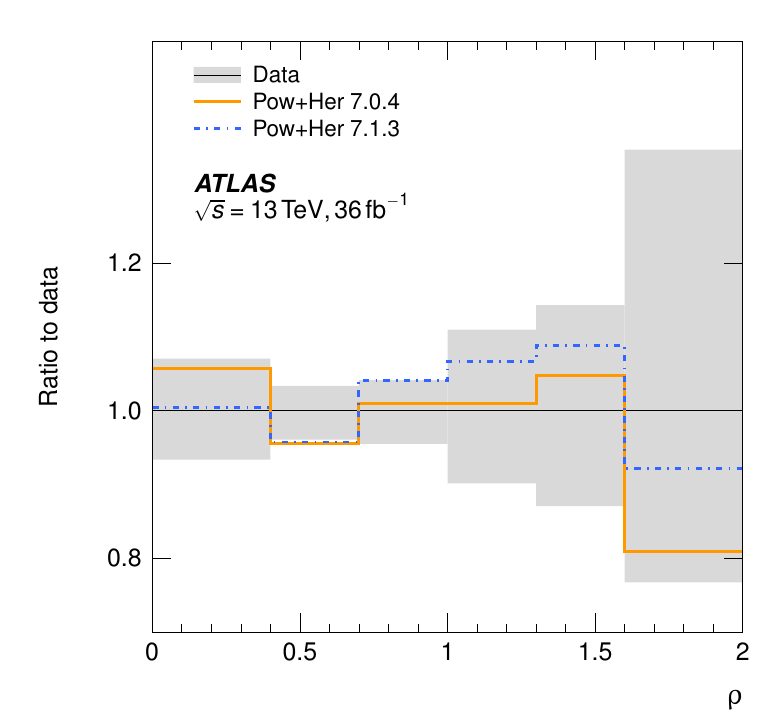}
}
\subfigure[][]{
\includegraphics[width=0.45\textwidth]{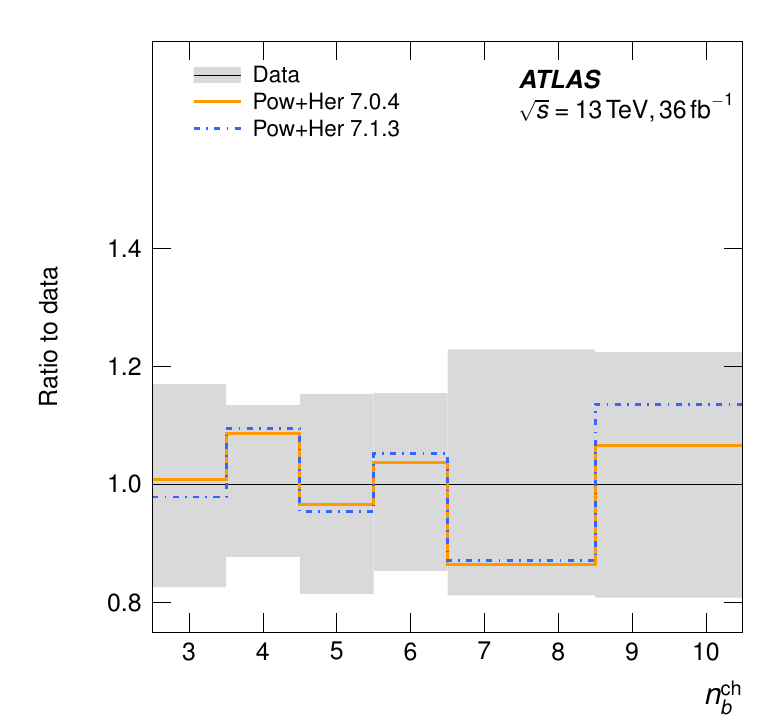}
}
\caption{Comparison of particle-level observables between
\POWHER[7] versions and unfolded data for (a) \zblc, (b) \zbtc, (c) $\rho$,
and (d) \nsvtrk.
} \label{fig:herwig}
\end{figure}

\begin{figure}[h!]
\centering
 
\subfigure[][]{
\includegraphics[width=0.45\textwidth]{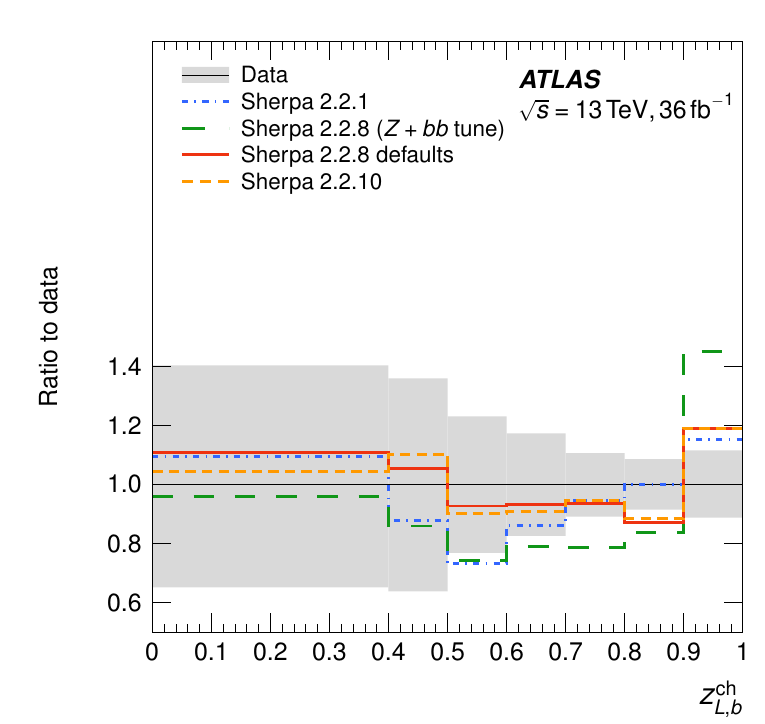}
}
\subfigure[][]{
\includegraphics[width=0.45\textwidth]{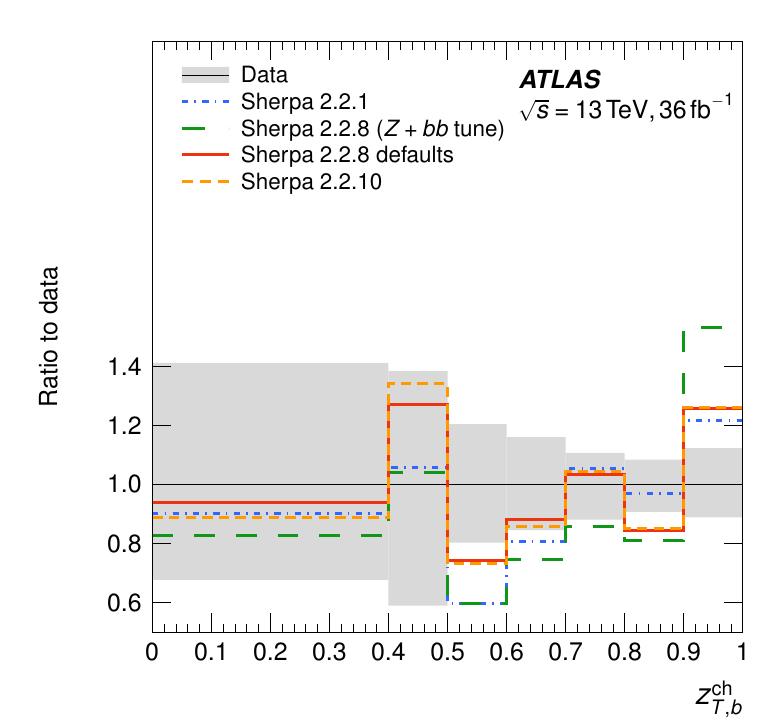}
} \\
\subfigure[][]{
\includegraphics[width=0.45\textwidth]{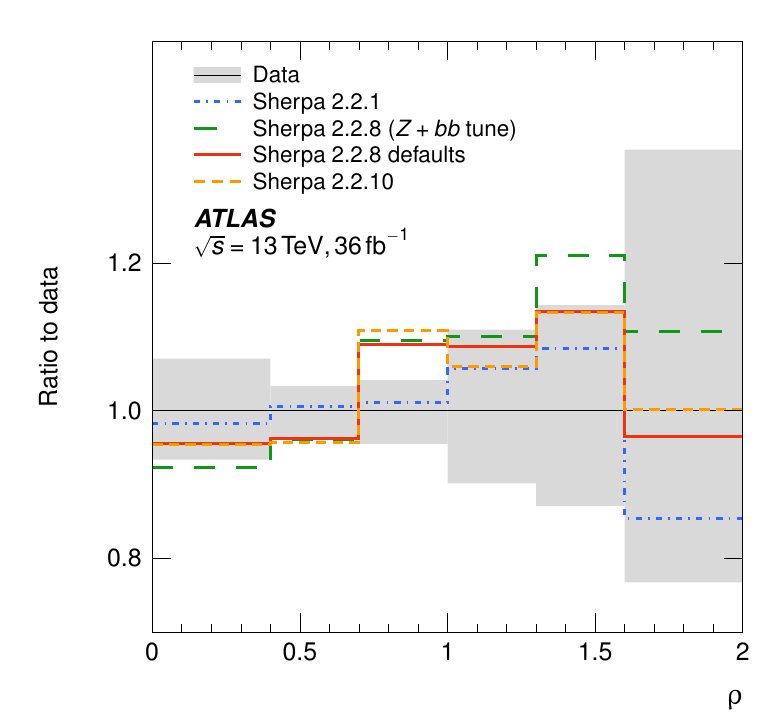}
}
\subfigure[][]{
\includegraphics[width=0.45\textwidth]{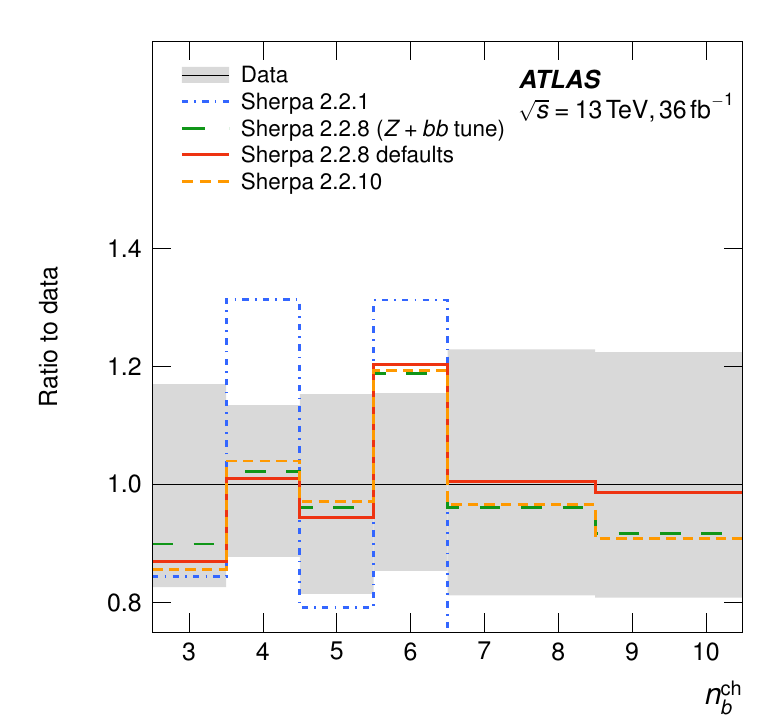}
}
\caption{Comparison of particle-level observables between \SHERPA
versions and unfolded data for (a) \zblc, (b) \zbtc, (c) $\rho$, and (d)
\nsvtrk.
} \label{fig:sherpa}
\end{figure}

Comparisons with \POWHER[7] and \SHERPA predictions are shown in
Figures~\ref{fig:herwig} and~\ref{fig:sherpa}, respectively.
While \HERWIG[7.0.4] appears to overestimate $b$-hadrons with soft fragmentation,
developments introduced in version~7.1 give significantly better agreement
with LHC data.
There has also been substantial progress in recent versions of \SHERPA: the
choice of parameter values used by ATLAS in \SHERPA[2.2.1] showed significant
mismodeling of \nsvtrk, but in versions~2.2.8 and 2.2.10 this is largely
corrected.
Two \SHERPA[2.2.8] curves are shown: in the first curve the default settings are used,
and in the second the settings recommended in Ref.~\cite{Hoche:2019ncc} are
used.
In the second, the shower evolution variable in the splitting functions
involving heavy-flavor quarks was tuned according to LHC $Z+b\bar{b}$ production
measurements.
The prediction from the default settings is in much better agreement with the data.
 
Corresponding $p$-values are calculated by approximating the posterior
probability distribution over the parameters of interest as a multivariate normal distribution and
taking the log-likelihood of a given prediction as the test statistic.
The $p$-values of the comparisons between unfolded data and generator predictions
are presented in Table~\ref{tab:pvalues}.
\POWPY[8] A14 and its variations predict the data reasonably well, as do
aMC@NLO+\PYTHIA[8], \POWPY[8] Monash, \HERWIG[7.1.3], and \SHERPA[2.2.8] and 2.2.10.
\SHERPA[2.2.8] with modified $g \to b\bar{b}$ splitting functions and \POWPY[8] with
the Peterson fragmentation model are strongly
disfavored by the data for the $z$ observables, as is \SHERPA[2.1.1] for \nsvtrk.
\POWPY[8] A14 with $\asfsr = 0.111$ and \POWHER[7.0.4] are in mild disagreement with
the data for the $z$ observables and $\rho$.

\begin{table}
\centering
\caption{Summary of $p$-values for various MC generator configurations based on
the observed data distributions.}
\label{tab:pvalues}
\vspace{0.2cm}
\begin{tabular}{ l | c | c | c | c }
\toprule
Generator configuration & \zbtc\ $p$-value & \zblc\ $p$-value & $\rho$ $p$-value & \nsvtrk\ $p$-value \\
\midrule
\powpy\!8 A14                       & 0.24 & 0.85 & 0.19 & 0.98 \\
\powpy\!8 A14 $\asfsr = 0.139$      & 0.09 & 0.33 & 0.28 & 0.98 \\
\powpy\!8 A14 $\asfsr = 0.111$      & 0.04 & 0.32 & 0.07 & 0.98 \\
\powpy\!8 A14 $r_B = 1.05$          & 0.09 & 0.48 & 0.28 & 0.98 \\
\amcpy\!8 A14                       & 0.29 & 0.92 & 0.21 & 0.98 \\
\powpy\!8 Monash                    & 0.13 & 0.57 & 0.32 & 0.97 \\
\powpy\!8 Monash Peterson           & 0.01 & 0.02 & 0.10 & 0.96 \\
\POWHER[7.0.4]                      & 0.16 & 0.46 & 0.11 & 0.98 \\
\POWHER[7.1.3]                      & 0.23 & 0.71 & 0.16 & 0.96 \\
\SHERPA[2.2.1]                      & 0.08 & 0.42 & 0.53 & 0.01 \\
\SHERPA[2.2.8] ($Z + b\bar{b}$ tune)      & ~~~~0.0005 & ~~0.005 & 0.19 & 0.48 \\
\SHERPA[2.2.8]                      & 0.10 & 0.47 & 0.11 & 0.61 \\
\SHERPA[2.2.10]                     & 0.07 & 0.53 & 0.07 & 0.40 \\
\hline
\end{tabular}
\end{table}

\FloatBarrier
% End of text imported from the .//results.tex input file
 
% The next lines are included from the .//conclusion.tex input file
\section{Conclusion}
\label{sec:conclusion}

Due to their ubiquity in decays of top quarks and Higgs bosons,
$b$-quarks have a substantial role to play in the analysis of LHC collision data
and are likely to continue to be invaluable in future collider experiments.
It is therefore important that the evolution and hadronization of heavy quarks be
well-understood.
As such, measurements of several observables sensitive to the fragmentation of
$b$-quarks have been carried out with the ATLAS detector, using $b$-quarks from
top-quark decays in $\ttbar \to e\nu b\mu\nu b$ events in 36~fb${}^{-1}$ of LHC $pp$
collision data at $\sqrt{s} = 13$~\TeV.
They complement similar measurements from $e^+e^-$ collider experiments
in which the $b$-quarks originate from a color-singlet $Z/\gamma^*$, by
probing the fragmentation process in a new environment as well as introducing
new observables that are sensitive to the local hadronic energy flow.
Since they provide a unique test of heavy-quark-fragmentation modeling at the
LHC, the measurements were corrected for detector effects and compared with
state-of-the-art MC event generator predictions.
With a few exceptions (e.g. \powpy\!8 Monash Peterson, \powpy\!8 A14 $\asfsr =
0.111$, and \SHERPA[2.2.8] tuned to $Z + b\bar{b}$ measurements), generators
tuned to a combination of lepton- and hadron-collider measurements yield
predictions that are found to agree with the observed LHC data.
% End of text imported from the .//conclusion.tex input file
 
\clearpage
 
\section*{Acknowledgments}
 
% The next lines are included from the .//acknowledgements/Acknowledgements.tex input file

We thank CERN for the very successful operation of the LHC, as well as the
support staff from our institutions without whom ATLAS could not be
operated efficiently.
 
We acknowledge the support of
ANPCyT, Argentina;
YerPhI, Armenia;
ARC, Australia;
BMWFW and FWF, Austria;
ANAS, Azerbaijan;
SSTC, Belarus;
CNPq and FAPESP, Brazil;
NSERC, NRC and CFI, Canada;
CERN;
ANID, Chile;
CAS, MOST and NSFC, China;
Minciencias, Colombia;
MEYS CR, Czech Republic;
DNRF and DNSRC, Denmark;
IN2P3-CNRS and CEA-DRF/IRFU, France;
SRNSFG, Georgia;
BMBF, HGF and MPG, Germany;
GSRI, Greece;
RGC and Hong Kong SAR, China;
ISF and Benoziyo Center, Israel;
INFN, Italy;
MEXT and JSPS, Japan;
CNRST, Morocco;
NWO, Netherlands;
RCN, Norway;
MEiN, Poland;
FCT, Portugal;
MNE/IFA, Romania;
JINR;
MES of Russia and NRC KI, Russian Federation;
MESTD, Serbia;
MSSR, Slovakia;
ARRS and MIZ\v{S}, Slovenia;
DSI/NRF, South Africa;
MICINN, Spain;
SRC and Wallenberg Foundation, Sweden;
SERI, SNSF and Cantons of Bern and Geneva, Switzerland;
MOST, Taiwan;
TAEK, Turkey;
STFC, United Kingdom;
DOE and NSF, United States of America.
In addition, individual groups and members have received support from
BCKDF, CANARIE, Compute Canada and CRC, Canada;
COST, ERC, ERDF, Horizon 2020 and Marie Sk{\l}odowska-Curie Actions, European Union;
Investissements d'Avenir Labex, Investissements d'Avenir Idex and ANR, France;
DFG and AvH Foundation, Germany;
Herakleitos, Thales and Aristeia programmes co-financed by EU-ESF and the Greek NSRF, Greece;
BSF-NSF and GIF, Israel;
Norwegian Financial Mechanism 2014-2021, Norway;
NCN and NAWA, Poland;
La Caixa Banking Foundation, CERCA Programme Generalitat de Catalunya and PROMETEO and GenT Programmes Generalitat Valenciana, Spain;
G\"{o}ran Gustafssons Stiftelse, Sweden;
The Royal Society and Leverhulme Trust, United Kingdom.
 
The crucial computing support from all WLCG partners is acknowledged gratefully, in particular from CERN, the ATLAS Tier-1 facilities at TRIUMF (Canada), NDGF (Denmark, Norway, Sweden), CC-IN2P3 (France), KIT/GridKA (Germany), INFN-CNAF (Italy), NL-T1 (Netherlands), PIC (Spain), ASGC (Taiwan), RAL (UK) and BNL (USA), the Tier-2 facilities worldwide and large non-WLCG resource providers. Major contributors of computing resources are listed in Ref.~\cite{ATL-SOFT-PUB-2021-003}.
 
% End of text imported from the .//acknowledgements/Acknowledgements.tex input file

\clearpage

\printbibliography

\clearpage
 
\clearpage
\input{atlas_authlist}

\end{document}

%% file: atlas_authlist.tex
% ATLAS Collaboration author list
% Reference date of TOPQ-2017-19 is 2020-08-25
% Author list last updated on date 10-SEP-22
% Data extracted on 10-Sep-2022 for paper reference TOPQ-2017-19
% at 6:08pm
 
\begin{flushleft}
\hypersetup{urlcolor=black}
{\Large The ATLAS Collaboration}

\bigskip

\AtlasOrcid[0000-0002-6665-4934]{G.~Aad}$^\textrm{\scriptsize 102}$,    
\AtlasOrcid[0000-0002-5888-2734]{B.~Abbott}$^\textrm{\scriptsize 128}$,    
\AtlasOrcid[0000-0002-7248-3203]{D.C.~Abbott}$^\textrm{\scriptsize 103}$,    
\AtlasOrcid[0000-0002-2788-3822]{A.~Abed~Abud}$^\textrm{\scriptsize 36}$,    
\AtlasOrcid[0000-0002-1002-1652]{K.~Abeling}$^\textrm{\scriptsize 53}$,    
\AtlasOrcid[0000-0002-2987-4006]{D.K.~Abhayasinghe}$^\textrm{\scriptsize 94}$,    
\AtlasOrcid[0000-0002-8496-9294]{S.H.~Abidi}$^\textrm{\scriptsize 29}$,    
\AtlasOrcid[0000-0002-8279-9324]{O.S.~AbouZeid}$^\textrm{\scriptsize 40}$,    
\AtlasOrcid{N.L.~Abraham}$^\textrm{\scriptsize 156}$,    
\AtlasOrcid[0000-0001-5329-6640]{H.~Abramowicz}$^\textrm{\scriptsize 161}$
% ...
\bigskip
\\

% ...
$^{29}$Physics Department, Brookhaven National Laboratory, Upton NY; United States of America.\\
% ...
$^{36}$CERN, Geneva; Switzerland.\\
% ...
$^{40}$Niels Bohr Institute, University of Copenhagen, Copenhagen; Denmark.\\
% ...
$^{53}$II. Physikalisches Institut, Georg-August-Universit\"{a}t G\"ottingen, G\"ottingen; Germany.\\
% ...
$^{94}$Department of Physics, Royal Holloway University of London, Egham; United Kingdom.\\
% ...
$^{102}$CPPM, Aix-Marseille Universit\'e, CNRS/IN2P3, Marseille; France.\\
$^{103}$Department of Physics, University of Massachusetts, Amherst MA; United States of America.\\
% ...
$^{128}$Homer L. Dodge Department of Physics and Astronomy, University of Oklahoma, Norman OK; United States of America.\\
% ...
$^{156}$Department of Physics and Astronomy, University of Sussex, Brighton; United Kingdom.\\
% ...
$^{161}$Raymond and Beverly Sackler School of Physics and Astronomy, Tel Aviv University, Tel Aviv; Israel.\\
% ...

\end{flushleft}

% Created with Glance <Atlas.Glance@cern.ch>